\documentclass[ALICE,manyauthors]{cernphprep}

\usepackage{hyperref}
\usepackage{lineno}

\usepackage[english]{babel}
\usepackage{graphicx}
\usepackage{verbatim}
\usepackage{amsfonts}
\usepackage{makeidx}
\usepackage{color}
\usepackage{amsmath}
\usepackage{epsfig}
\usepackage{epstopdf}
\usepackage{hyperref}
\usepackage{cite}

\usepackage{subfigure}
\usepackage{rotating}
\usepackage{lscape}

\usepackage[comma,square,numbers,sort&compress]{natbib}

\newcommand{\pp}{\text{pp}}
\newcommand{\ppbar}{\text{p}\overline{\text{p}}}

\newcommand{\pPb}{\mbox{p--Pb}}
\newcommand{\sqrts}{\sqrt{s}}
\newcommand{\sqrtsNN}{\sqrt{s_{\rm NN}}}

\newcommand{\TeV}{\mathrm{TeV}}

\newcommand{\kev}{\mathrm{keV}}
\newcommand{\mev}{\mathrm{MeV}}

\newcommand{\mevcc}{\mathrm{MeV}/c^2}
\newcommand{\gev}{\mathrm{GeV}}
\newcommand{\gevc}{\mathrm{GeV}/c}

\newcommand{\cm}{\mathrm{cm}}

\newcommand{\mum}{\mathrm{\mu m}}

\newcommand{\pt}{p_{\rm T}}

\newcommand{\DtoKpi}{{\rm D}^0 \to {\rm K}^-\pi^+}
\newcommand{\DtoKpipi}{{\rm D}^+\to {\rm K}^-\pi^+\pi^+}
\newcommand{\DstartoDpi}{{\rm D}^{*+} \to {\rm D}^0 \pi^+}

\newcommand{\Dzero}{{\rm D^0}}
\newcommand{\Dzerobar}{\overline{{\rm D^0}}}
\newcommand{\Dstar}{{\rm D^{*+}}}

\newcommand{\Dplus}{{\rm D^+}}

\newcommand{\Ds}{{\rm D^+_{\rm s}}}
\newcommand{\Lc}{{\rm \Lambda^+_{\rm c}}}
\newcommand{\Jpsi}{{\rm J}/\psi}
\newcommand{\ccbar}{${\rm c}\bar{\rm c}$~}
\newcommand{\bbbar}{${\rm b}\bar{\rm b}$~}
\newcommand{\nbinv}{{\rm nb^{-1}}}

\newcommand{\dEdx}{{\rm d}E/{\rm d}x}

\newcommand{\Ntrk}{N_{\rm tracklets}}
\newcommand{\Nvzero}{N_{\rm V0}}
\newcommand{\dNdEta}{{\rm d}N_{\rm ch}/{\rm d}\eta}

\newcommand{\dNDzdydpt}{{\rm d}^2N^{\rm \Dzero}/{\rm d}y {\rm d}\pt}
\newcommand{\fB}{f_{\rm B}}

\begin{document}%

\begin{titlepage}
\PHyear{2015}
\PHnumber{091}      
\PHdate{30 March}  
%


\title{Measurement of charm and beauty production at central rapidity versus charged-particle multiplicity in proton--proton collisions at ${\mathbf \sqrts=7}$~TeV}
\ShortTitle{Charm and beauty production versus charged-particle multiplicity}   

\Collaboration{ALICE Collaboration\thanks{See Appendix~\ref{app:collab} for the list of collaboration members}}
\ShortAuthor{ALICE Collaboration} 

\begin{abstract}
Prompt D meson and non-prompt $\Jpsi$ yields are studied as a function of the multiplicity of charged particles produced in inelastic proton--proton collisions at a centre-of-mass energy of $\sqrts=7$~TeV. The results are reported as a ratio between yields in a given multiplicity interval normalised to the multiplicity-integrated ones (relative yields). They are shown as a function of the multiplicity of charged particles normalised to the average value for inelastic collisions (relative charged-particle multiplicity). 
$\Dzero$, $\Dplus$ and $\Dstar$ mesons are measured in five $\pt$ intervals from $1~\gevc$ to $20~\gevc$ and for $|y|<0.5$ via their hadronic decays. The D-meson relative yield is found to increase with increasing charged-particle multiplicity. For events with multiplicity six times higher than the average multiplicity of inelastic collisions, a yield enhancement of a factor about 15 relative to the multiplicity-integrated yield in inelastic collisions is observed. 
The yield enhancement is independent of transverse momentum within the uncertainties of the measurement. 
The $\Dzero$-meson relative yield is also measured as a function of the relative multiplicity at forward pseudo-rapidity.  
The non-prompt $\Jpsi$, i.e. the B hadron, contribution to the inclusive $\Jpsi$ production is measured in the di-electron decay channel at central rapidity. It is evaluated for $\pt>1.3~\gevc$ and $|y|<0.9$, and extrapolated to $\pt>0$. The fraction of non-prompt $\Jpsi$ in the inclusive $\Jpsi$ yields shows no dependence on the charged-particle multiplicity at central rapidity. 
Charm and beauty hadron relative yields exhibit a similar increase with increasing charged-particle multiplicity. 
The measurements are compared to PYTHIA~8, EPOS~3 and percolation calculations. 
\end{abstract}
\end{titlepage}
\setcounter{page}{2}

%
%

\section{Introduction}
\label{sec:intro}
The study of the production of hadrons containing heavy quarks, i.e.\ charm and 
beauty, in proton--proton ($\pp$) collisions at the Large Hadron Collider (LHC)
provides a way to test calculations based on perturbative Quantum 
Chromodynamics (pQCD) at the highest available collision energies. 
The inclusive production cross sections of charm mesons measured in pp 
collisions at the LHC at both central~\cite{ALICE:2011aa,Abelev:2012vra} and 
forward~\cite{Aaij:2013mga} rapidity are described by 
theoretical predictions based on pQCD calculations with the collinear 
factorisation approach at next-to-leading order (e.g. in the general-mass 
variable-flavour-number scheme, GM-VFNS~\cite{Kniehl:2012ti}) or at fixed 
order with next-to-leading-log resummation (FONLL~\cite{Cacciari:1998it,Cacciari:2001td,Cacciari:2012ny,Kniehl:2005ej}) within theoretical uncertainties.
The comparisons suggest that charm production is under (over) estimated by the central values of the FONLL (GM-VFNS) calculations.
The measured D-meson production cross sections in pp collisions at the LHC can 
also be described by pQCD calculations performed in the framework of 
$k_{\rm T}$-factorisation in the leading order (LO) 
approximation~\cite{Maciula:2013wg}.
Beauty production cross section measurements in pp collisions at 
$\sqrt{s} = 7$~TeV~\cite{Abelev:2012gx,Abelev:2012sca,ATLAS:2013cia,Khachatryan:2010yr,Aaij:2010gn} are well described by  implementations of FONLL and GM-VFNS~\cite{Cacciari:2012ny,Kniehl:2011bk}.
In the case of B mesons, the measured cross sections are close to the central
value of the FONLL and GM-VFNS predictions.
A similar situation was observed in $\ppbar$~collisions 
at $\sqrt{s} = 1.96$~TeV at the FNAL Tevatron collider~\cite{Acosta:2004yw,Cacciari:2003uh,Kniehl:2008zza}.

The measurement of heavy-flavour production in $\pp$~collisions as 
a function of the charged-particle multiplicity produced in the collision 
could provide insight into the processes occurring in the collision at the 
partonic level and the interplay between the hard and soft mechanisms in 
particle production.
These aspects are expected to depend on the energy and on the 
impact parameter (the distance between the colliding protons in the plane
perpendicular to the beam direction) of the pp 
collision~\cite{Frankfurt:2003td,Frankfurt:2010ea,Azarkin:2014cja}.
In the impact parameter representation of proton–proton collisions, 
the overlap of the nucleon wave functions in proton--proton collisions can be 
described by a geometrical picture with two separate transverse distance 
scales: the impact parameter of the collision and the transverse spatial partonic distribution~\cite{Gribov1963107,Block:1984ru,Frankfurt:2010ea,Dremin2013241}.%
In particular, pp collisions with a hard parton-parton scattering are
predicted to be more central (i.e.\ have smaller impact parameter) than 
minimum-bias events~\cite{Frankfurt:2010ea,Strikman:2011zz}.

The NA27 Collaboration observed in 1988 that the average charged-particle 
multiplicity in events with open charm production was higher by about $20\%$ than in 
events without charm production~\cite{AguilarBenitez:1988js}. 
A softening of the momentum spectra of hadrons produced in association with 
charm was also observed. 
This result was interpreted as a consequence of the more central nature
of collisions leading to charm production.

At LHC energies, two additional contributions to charm production and its relation to multiplicity have to be considered. 
The first effect is the likely larger amount of gluon radiation associated to 
the short distance production processes at larger energies and particle transverse 
momenta.
The second is the contribution of Multiple-Parton Interactions 
(MPI)~\cite{Bartalini:2010su,Sjostrand:1987su,Porteboeuf:2010dw}, i.e.\ several 
hard partonic interactions occurring in a single $\pp$~collision.
In this context, pQCD-inspired models describe the final-state particles 
produced in hadronic collisions with a two-component approach, namely an
initial hard partonic scattering process, that gives rise to collimated 
clusters of hadrons (jets), and an underlying event, consisting of the 
final-state particles that are not associated with the initial hard scattering.
While the hard scattering process can be computed with a pQCD approach, the 
description of the underlying event, which is thought to be dominated by
particles produced in soft processes and by perturbative (mini)jets with 
relatively small transverse momenta (soft MPIs), is based on a
phenomenological model.
In particular, pQCD-based models of MPIs 
provide a consistent way to describe high multiplicity pp collisions, and 
have been implemented in recent Monte Carlo generators like 
PYTHIA~6~\cite{Sjostrand:2006za}, PYTHIA~8~\cite{Sjostrand:2007gs}, 
and HERWIG~\cite{Bahr:2008pv}.
Measurements by the CMS Collaboration of jet and underlying 
event properties as a function of multiplicity in pp collisions at 
$\sqrt{s}$ = 7 TeV can be better described by event generators including 
MPI~\cite{Chatrchyan:2013ala,ALICE:2011ac}.
The analysis of minijet production performed by the ALICE 
Collaboration~\cite{Abelev:2013sqa} indicates that high multiplicities in pp 
collisions are reached through a high number of MPIs and a higher than average 
number of fragments per parton.
Upward fluctuations of the gluon density in the colliding protons are also 
advocated to describe the results from high multiplicity pp collisions at the 
LHC~\cite{Azarkin:2014cja,Strikman:2011ar,Kopeliovich:2013yfa}.
Indeed, the transverse structure of the proton, as probed in hard partonic
scattering processes, is predicted to play a crucial role in defining the underlying event structure and the probability of MPIs~\cite{Strikman:2011zz}.
In the heavy-flavour sector, the LHCb Collaboration reported measurements of 
double charm production in pp collisions at the LHC ($\Dzero+X$, $\Jpsi+X$ and $\Jpsi+\Jpsi$ where $X=\Dzero, \, \Dplus, \, \Ds, \, \Lc$), which suggest that MPIs also play a role at the hard momentum scale relevant
for c$\overline{\mathrm c}$ production~\cite{Aaij:2012dz,Aaij:2011yc}. 

The ALICE Collaboration published the first measurement of inclusive $\Jpsi$ production as a function of charged-particle multiplicity, expressed as the pseudo-rapidity density of charged particles $\dNdEta$ at mid-rapidity, in $\pp$~collisions at $\sqrt{s} = 7$~TeV~\cite{Abelev:2012rz}. 
An approximately linear increase of the yield of $\Jpsi$ with the charged-particle multiplicity was observed in a multiplicity range reaching four times the average multiplicity $\langle \dNdEta \rangle$.
The measurements at $|y|<0.9$ and $2.5<y<4.0$ were compatible within the uncertainties.
Both the larger amount of gluon radiation and the contribution of MPI in collisions where heavy quarks are produced can induce 
a correlation between the yield of quarkonia and the charged-particle multiplicity produced in the collision.
The measured rise of $\Jpsi$ yield with increasing multiplicity can also
be described in the framework of string interaction or parton saturation 
models. 
In particular, in Ref.~\cite{Ferreiro:2012fb} a stronger--than--linear trend in the high density domain is anticipated as a consequence 
of the interaction (overlap) of strings, which reduces the 
effective number of sources for soft-particle production.
The increasing trend of $\Jpsi$ yield with multiplicity is also described 
in a framework in which high multiplicities are attained in pp collisions due 
to the contribution of higher Fock states in the proton, leading to a larger 
number of gluons participating in the collision~\cite{Kopeliovich:2013yfa}.

It is also worth pointing out that the charged-particle densities attained in high-multiplicity $\pp$~collisions at the LHC are of the same order of magnitude as those measured in semi-peripheral heavy-ion collisions at lower centre-of-mass energies~\cite{Alver:2010ck}. 
In those heavy-ion collisions, the measured momentum distributions of light hadrons indicate that the system undergoes a collective expansion, which can be described in terms of hydrodynamics. 
Recent measurements in high-multiplicity $\pPb$~collisions at $\sqrtsNN=5.02$~TeV~\cite{ABELEV:2013wsa,Abelev:2012ola,CMS:2012qk,Chatrchyan:2013nka,tagkey201360,Aad:2012gla} and in high-multiplicity $\pp$~collisions at the LHC~\cite{Khachatryan:2010gv} indicate that such a collective behaviour could also be at play in these systems. 
If charm quarks were to follow a collective motion in high-multiplicity events, their momentum spectra would be altered, and the heavy-flavour hadron relative yields at high multiplicity would vary as a function of $\pt$~\cite{Vogel:2013wqa}.

The measurements of the $\pt$-differential prompt D meson and non-prompt $\Jpsi$ cross sections in $\pp$~collisions at $\sqrts=7$~TeV with the ALICE experiment at the LHC were published in references~\cite{ALICE:2011aa,Abelev:2012gx}. 
In this paper, we report the measurement of the relative open heavy-flavour production yields as a function of the charged-particle multiplicity in $\pp$~collisions at $\sqrt{s}=7$~TeV. 
Open charm and beauty production is measured by reconstructing prompt D mesons
and non-prompt $\Jpsi$, i.e.\ $\Jpsi$ mesons coming from the decay of beauty 
hadrons. 
The experimental setup and the multiplicity estimation are described in Secs.~\ref{sec:detector} and~\ref{sec:Multiplicity}, respectively. 
Prompt $\Dzero$, $\Dplus$, $\Dstar$ mesons were measured at central rapidity, $|y|<0.5$, in six multiplicity intervals and five $\pt$ intervals from 1 $\gevc$ to 20 $\gevc$ (Sec.~\ref{sec:Dmeson}). The non-prompt fraction of $\Jpsi$ production was measured in the rapidity interval $|y|<0.9$ in five multiplicity intervals and for $\pt>1.3~\gevc$ and extrapolated to $\pt>0$ (Sec.~\ref{sec:Jpsi}). 
The relative yields as a function of charged-particle multiplicity are compared in Sec.~\ref{sec:results}. 
Finally, model calculations are discussed and compared with data in Sec.~\ref{sec:MCcalculations}.

\section{Experimental apparatus and data sample}
\label{sec:detector}
The ALICE apparatus~\cite{Aamodt:2008zz} consists of a central barrel detector covering the pseudo-rapidity interval $|\eta|<0.9$, a forward muon spectrometer covering the pseudo-rapidity interval $-4.0<\eta<-2.5$, and a set of detectors at forward and backward rapidities used for triggering and event characterization. In the following, the subsystems that are relevant for the D meson and non-prompt $\Jpsi$ analyses are described. 

The central barrel detectors are located inside a large solenoidal magnet, which provides a magnetic field of 0.5 T along the beam direction ($z$ axis in the ALICE reference frame). 
Tracking and particle identification are performed using the information provided by the Inner Tracking System (ITS), the Time Projection Chamber (TPC) and the Time Of Flight (TOF) detectors, that have full azimuthal coverage in the pseudo-rapidity interval $|\eta|<0.9$. 
The detector closest to the beam axis is the ITS, which is composed of six cylindrical layers of silicon detectors, with radial distances from the beam axis ranging from 3.9~cm to 43.0~cm. 
The two innermost layers, with average radii of 3.9~cm and 7.6~cm, are equipped with Silicon Pixel Detectors (SPD). 
The two SPD layers, covering the pseudo-rapidity ranges of $|\eta|< 2.0$ and $|\eta|< 1.4$ respectively, have 1200 SPD readout chips.  
The two intermediate layers are made of Silicon Drift Detectors (SDD), while Silicon Strip Detectors (SSD) equip the two outermost layers. 
The high spatial resolution of the silicon sensors, together with the low material budget (on average 7.7\% of a radiation length for tracks crossing the ITS perpendicularly to the detector surfaces, i.e.\ $\eta=0$) and the small distance of the innermost layer from the beam vacuum tube, allow for the measurement of the track impact parameter in the transverse plane ($d_0$), i.e.\ the distance of closest approach of the track to the primary vertex in the plane transverse to the beam direction, with a resolution better than 75~$\mu$m for transverse momenta $\pt>1~\gevc$~\cite{Aamodt:2010aa}.
The SPD provides also a measurement of the multiplicity of charged particles produced in the collision based on track segments (tracklets) built by associating pairs of hits in the two SPD layers.

At larger radii ($85<r<247~\cm$), a 510 cm long cylindrical TPC~\cite{Alme2010316} provides track reconstruction with up to 159 three-dimensional space points per track, as well as particle identification via the measurement of the specific energy deposit $\dEdx$ in the gas.
The charged particle identification capability of the TPC is supplemented by the TOF~\cite{Abelev:2014ffa}, which is equipped with Multi-gap Resistive Plate Chambers  (MRPCs) located at radial distances between 377 and 399 cm from the beam axis. The overall TOF resolution including the uncertainty on the time at which the collision took place, and the tracking and momentum resolution was about 160~ps for the data-taking period considered in these analyses. 

The V0 detector~\cite{Abbas:2013taa}, used for triggering and for estimating 
the multiplicity of charged particles in the forward rapidity region, 
consists of two arrays of 32 scintillators each, placed around the beam vacuum tube on either side of the interaction region at $z =-90$ cm and $z=+340$ cm. The two arrays cover the pseudo-rapidity intervals  $-3.7 < \eta < -1.7$ and $2.8 < \eta < 5.1$, respectively.

The data from proton--proton (pp) collisions at a centre-of-mass energy of $\sqrt{s}=7~\TeV$ used for the analyses were recorded in 2010. 
The data sample consists of about 314 million minimum-bias (MB) events, corresponding to an integrated luminosity of $\mathcal{L}_{\rm int} \simeq 5~\nbinv$. 
Minimum-bias collisions were triggered by requiring at least one hit in either of the V0 counters or in the SPD ($|\eta|<2$), in coincidence with the arrival time of proton bunches from both directions. 
This trigger was estimated to be sensitive to about 85\% of the pp inelastic cross section~\cite{Abelev:2012sea}.

To enrich the data sample with high multiplicity events, a High Multiplicity (HM) trigger based on the multiplicity information provided by the outer SPD layer was also used. Each readout chip of the SPD promptly asserts a digital pulse, called FastOR bit, on the presence of at least one firing pixel.
A sample of about 6 million events was collected applying a selection on the minimum number of readout chips
having asserted this digital pulse.
The threshold was configured to select the $\approx 0.7\%$ of the events with highest number of hits in the outer SPD layer.
This HM-trigger sample ($\mathcal{L}_{\rm int} \simeq 14~\nbinv$) provides an increase of statistics by a factor of about 2.8 relative to the MB trigger 
for events with more than 50 tracklets, corresponding to about six times the average multiplicity.

Only events with interaction vertex reconstructed from tracks with a coordinate $|z|<10$~cm along the beam line were used for the analysis.
In the considered data samples, the instantaneous luminosity was limited to 0.6--$1.2\times 10^{29}~\mathrm{cm}^{-2}\mathrm{s}^{-1}$ by displacing the beams in the transverse plane by 3.8 times the RMS of their transverse profile. 
In this way, the interaction probability per bunch crossing was kept in the range 0.04--0.08, with a probability of collision pile-up below 4\% per triggered event.
An algorithm to detect multiple interaction vertices based on SPD track segments, or tracklets, was used to further reduce the pile-up contribution. 
An event is rejected from the analysed data sample if a second interaction vertex is found, which has at least three associated tracklets, and is separated from the first one by more than 0.8 cm along $z$. This removes about $48\%$ of the pile-up events.
The remaining pile-up contamination has two contributions: events with pile-up of collisions with $\Delta z < 0.8$~cm and events in which
the piled-up collisions have low-multiplicity (less than three charged particles reconstructed in the SPD). 
In the case of pile-up of collisions with small separation along $z$, the multiplicity estimation may be biased because some of the tracklets of charged particles from different interactions may be added together. 
According to simulations, the number of tracklets results to be biased when the piled-up vertices are separated along $z$ by less than 0.6~cm. 
Combining this result with the shape of the luminous region along the beam direction and the maximum pile-up rate of $4\%$, the overall probability that two piled-up events induce a bias in the determination of multiplicity was found to be lower than $0.3\%$. 
The fraction of events with biased number of tracklets increases with increasing multiplicity and it was estimated to be below $2\%$ at the highest multiplicities considered in this analysis, while the resulting bias on the measured number of tracklets was found to be negligible in all the multiplicity classes.

\section{Multiplicity definition and corrections}
\label{sec:Multiplicity}
In the present analysis, the experimental estimator of the charged-particle multiplicity is the number of tracklets in the interval $|\eta|<1.0$ ($\Ntrk$). Tracklets are track segments defined by combining the clusters in the SPD detector with the reconstructed primary vertex position. Tracklets are required to point to the primary interaction vertex within $\pm1$~cm in the transverse plane and $\pm3$~cm in the $z$ direction~\cite{Aamodt:2008zz,Aamodt:2010aa}. 
This multiplicity estimator is the same as was used in previous studies performed for inclusive $\Jpsi$ production~\cite{Abelev:2012rz}. 
Monte Carlo simulations of the detector response have shown that $\Ntrk$ is proportional to the pseudo-rapidity density of the generated charged primary particles, $\dNdEta$, within $2\%$. 
Primary particles are defined as prompt particles produced in the collision and all decay products, except products from weak decays of strange particles. 
The pseudo-rapidity coverage of the SPD detector changes with the position of the
interaction vertex along the beam line, $z_{\rm vtx}$, and with time due to the variation of the number of inactive channels. The detector response over the analysed data taking period is equalised by means of a data-based correction, which is applied on an event-by-event basis depending on $z_{\rm vtx}$ and time.

The measurements in the $\Ntrk \in$ [1,49] interval are performed using minimum-bias triggered data, while those in the [50,80] range exploit the SPD-based HM trigger described above. The HM trigger is fully efficient for events with $\Ntrk>65$. 
The number of events and the D-meson candidate invariant mass distributions were corrected for the HM trigger inefficiency in the $\Ntrk \in$ [50,65] range by means of a data-driven re-weighting procedure. 
The $\Ntrk$-dependent event weights were defined from the ratio of the measured distributions of the number of tracklets in the HM and minimum-bias trigger samples. 
The effect of this correction on the per-event raw yield was of about 2.5\%. 
The average $\dNdEta$ of events in the highest $\Ntrk$ interval was determined from the minimum-bias sample.

The analysis results are presented as a function of the relative charged-particle multiplicity at central rapidity, $(\dNdEta)^{j} \big/ \langle \dNdEta \rangle$, where $\langle \dNdEta \rangle = 6.01 \pm 0.01 \, {\rm(stat.)} \, ^{+0.20}_{-0.12} \, {\rm(syst.)}$ is measured in inelastic \pp~collisions at $\sqrts=7$~TeV with at least one charged particle in $|\eta|<1.0$~\cite{Aamodt:2010pp}. 
The relative quantities are used to minimise the experimental uncertainties and to facilitate the comparison with other measurements and models. 
The considered $\Ntrk$ intervals and the corresponding relative charged-particle multiplicity ranges are summarised in Table~\ref{tab:NchCorr}. 
The highest $\Ntrk$ interval considered in the analysis extends to a multiplicity of about 9 times the
$\langle \dNdEta \rangle$ of inelastic \pp~collisions and the average multiplicity of events in this
$\Ntrk$ interval is about six times the $\langle \dNdEta \rangle$.
The uncertainty on $(\dNdEta)^{j} \big/ \langle \dNdEta \rangle$ is $6\%$; it includes the influence of 
{\it (i)} the determination of the $\Ntrk$ to $\dNdEta$ proportionality factor, $2\%$, 
{\it (ii)} its possible deviation from linearity, $5\%$,
{\it (iii)} and the uncertainty on the measured $\langle \dNdEta \rangle$. 
\begin{table}[!htbp]
\begin{center}

\begin{tabular}{ccccc}
\hline
 $\Ntrk$ & $ (\dNdEta)^{j} $ & 
	$(\dNdEta )^{j}  \big/ \langle \dNdEta \rangle$ 
	& $N_{\rm events}^{\Dzero} / 10^6$	& $N_{\rm events}^{\Jpsi} / 10^6$  \\ \hline
  $[1,8]$ & \phantom{0}2.7 & $0.45^{+0.03}_{-0.03}$ & 155.1 & -- \\[1.01ex]
  $[4,8]$ & \phantom{0}3.8 & $0.63^{+0.04}_{-0.04}$ &  -- & 89.0 \\[1.01ex]
 $[9,13]$ & \phantom{0}7.1 & $1.18^{+0.07}_{-0.07}$ &  \phantom{0}46.2 & 50.5 \\[1.01ex]
 $[14,19]$ &10.7 		& $1.78^{+0.10}_{-0.11}$ & \phantom{0}32.0 & 35.5 \\[1.01ex]
 $[20,30]$ & 15.8		 & $2.63^{+0.15}_{-0.17}$ & \phantom{0}24.7 & 28.0 \\[1.01ex]
 $[31,49]$ & 24.1 		& $4.01^{+0.23}_{-0.25}$ & \phantom{00}7.9 & \phantom{0}9.5 \\[1.01ex]
 $[50,80]$ & 36.7 		& $6.11^{+0.35}_{-0.39}$ & \phantom{00}1.7 &  -- \\ [1.01ex]
\hline
\end{tabular}
\caption{Summary of the multiplicity intervals used for the analyses. The number of reconstructed tracklets $\Ntrk$, the average charged-particle multiplicity $(\dNdEta)^{j}$, and the relative charged-particle multiplicity $(\dNdEta)^{j} \big/ \langle \dNdEta \rangle$ are detailed. The number of events analysed in the various multiplicity ranges for both the D-meson and $\Jpsi$ analysis are reported. 
The number of events for the $\Ntrk$ interval $[50,80]$ are corrected for the high multiplicity trigger efficiency, as explained in Sec.~\ref{sec:Multiplicity}.
\label{tab:NchCorr}
}
\end{center}
\end{table}

The analysis of $\Dzero$ production is also carried out as a function of the charged-particle multiplicity in the regions $-3.7<\eta<-1.7$ and $2.8<\eta<5.1$, as measured with the charge collected by the V0 scintillator counters, $\Nvzero$, reported in units of the minimum-ionizing-particle charge. 
The motivation for studying the multiplicity dependence of charmed-meson production also with this estimator is that the event multiplicity and the D-meson yields are evaluated in different pseudorapidity ranges, reducing the effects of auto-correlations. 
In contrast, with the $\Ntrk$ estimator also the D-meson decay products and the charged particles produced in the fragmentation of the same charm quark are included in the multiplicity evaluation.
Monte Carlo simulations demonstrate that $\Nvzero$ is proportional to the charged-particle multiplicity in that pseudo-rapidity interval. 
In this paper we report $\Dzero$ relative yields as a function of the relative uncorrected multiplicity in the V0 detector, $\Nvzero\big/ \langle \Nvzero \rangle$ (see Sec.~\ref{sec:DresultsVzero}).

\section{D-meson analysis}
\label{sec:Dmeson}

\subsection{D-meson reconstruction}
Charm production was studied by reconstructing $\Dzero$, $\Dplus$ and $\Dstar$ mesons, and their antiparticles, via their hadronic decay channels $\DtoKpi$ (with branching ratio, BR, of $3.88 \pm 0.05\%$), $\DtoKpipi$ (BR of $9.13\pm0.19\%$), and $\DstartoDpi$ (BR of $67.7\pm0.05\%$) with $\DtoKpi$~\cite{Agashe:2014kda}. 
D-meson candidates were selected with the same strategy as described in~\cite{ALICE:2011aa}. 
The selection of $\Dzero$ and $\Dplus$ decays (weak decays with mean proper decay length $c\tau\approx 123$ and $312~\mum$, respectively~\cite{Agashe:2014kda}) was based on the reconstruction of secondary vertices separated by few hundred microns from the interaction point. 
In the case of the $\Dstar$ strong decay, the decay topology of the produced $\Dzero$ was reconstructed.  
$\Dzero$ and $\Dplus$ candidates were formed using pairs and triplets of tracks with the proper charge sign combination, $|\eta|<0.8$, $\pt>0.3~\gev/c$, at least 70 associated space points (out of a maximum of 159) with $\chi^2/{\rm ndf} < 2$ of the momentum fit in the TPC, and at least two hits (out of 6) in the ITS, of which at least one had to be in either of the two SPD layers.
$\Dstar$ candidates were formed by combining $\Dzero$ candidates with tracks with $\pt>80~\mev/c$ and at least 3 hits in the ITS, out of which at least one should be in the SPD.
The selection of tracks with $|\eta|<0.8$ limits the D-meson acceptance in rapidity. 
The acceptance drops steeply to zero for $|y|> 0.5$ at low $\pt$ and $|y|> 0.8$ at $\pt> 5~\gev/c$.
A $\pt$-dependent fiducial acceptance cut, $|y_{\rm D}|<y_{\rm fid}(\pt)$, was therefore applied on the D-meson rapidity.
The cut value, $y_{\rm fid}(\pt)$, increases from 0.5 to 0.8 in the transverse momentum range $0<\pt<5~\gev/c$ according to a second-order polynomial function and it takes a constant value of 0.8 for $\pt>5~\gev/c$. 
The selection of the decay topology was based on the displacement of the decay tracks from the interaction vertex, the separation between the secondary and primary vertices, and the pointing angle of the reconstructed D-meson momentum and its flight line from the primary to the secondary vertex. 
The selections were tuned such that a large statistical significance of the signal and a selection efficiency as high as possible were achieved, which resulted in cut values that depend on the D-meson $\pt$ and species~\cite{ALICE:2011aa}.
The same selections were used in all the multiplicity intervals in order to minimise the effect of efficiency corrections in the ratio of the yields.
Pion and kaon identification based on the TPC and TOF detectors were used to obtain a further reduction of the background.
Cuts in units of resolution (at $\pm3\,\sigma$) were applied around the expected mean values of energy deposit $\dEdx$ and time-of-flight.
Tracks without TOF signal were identified using only the TPC information. 
Tracks with incompatible TPC and TOF response were considered as non-identified and were used in the analysis as both pion and kaon candidates.
Particle identification (PID) was not applied to the pion tracks from the $\Dstar$ decay.
This selection guarantees a reduction of the background by a factor of about 2 to 3 at low $\pt$, while preserving about $95\%$ of the signal.

\begin{figure}[tb!p]
\begin{center}
\includegraphics[width=0.85\columnwidth]{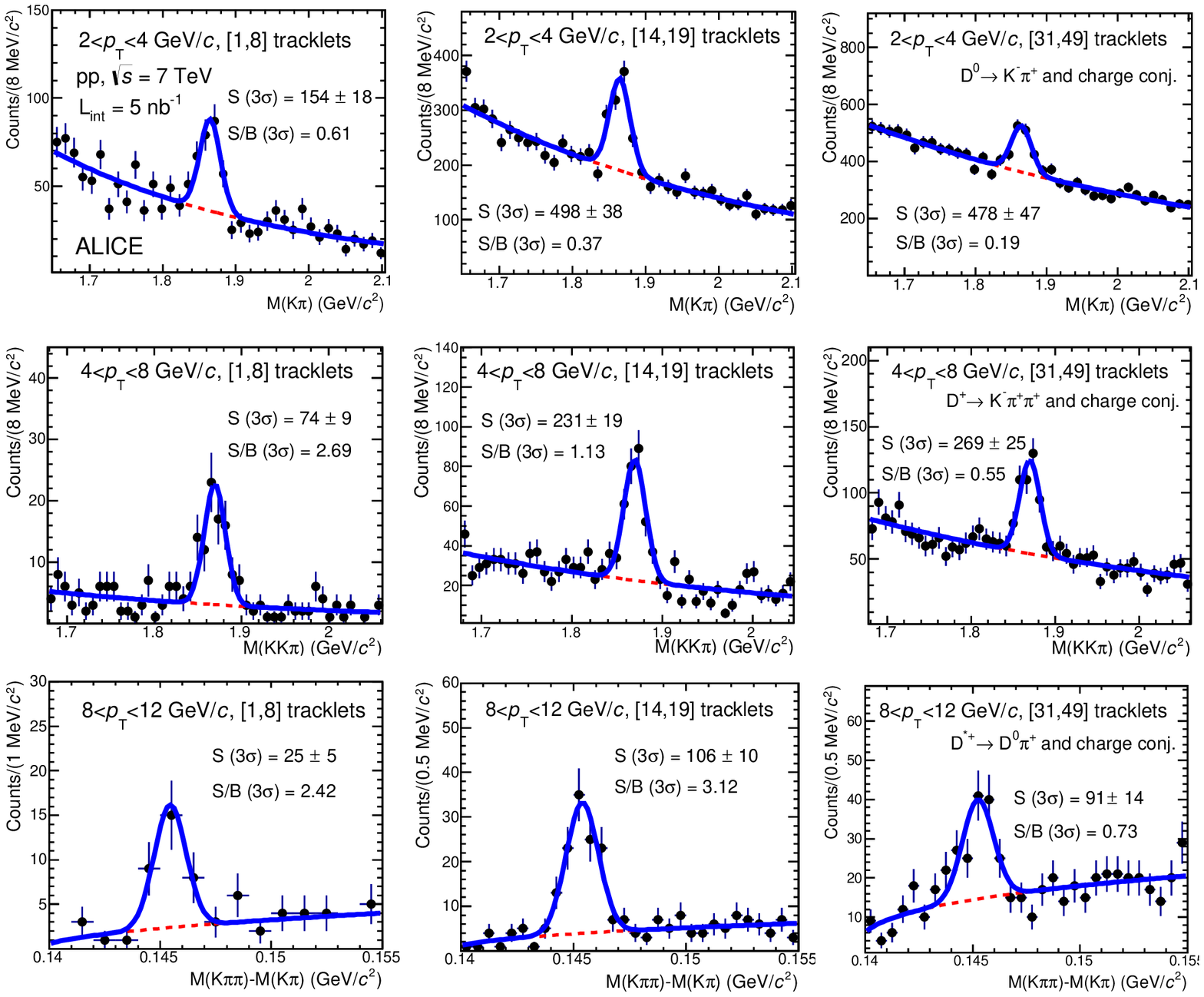}
\caption{
$\Dzero$ and $\Dplus$ invariant mass and $\Dstar$ mass difference distributions for selected $\pt$ and $\Ntrk$ intervals for \pp~collisions at~$\sqrt{s}=7$~TeV with $\mathcal{L}_{\rm int} = 5~\nbinv$. 
The $\Dzero$ distributions are shown for $2<\pt<4~\gevc$ (top-panels), the $\Dplus$ for $4<\pt<8~\gevc$ (middle-panels), and the $\Dstar$ for $8<\pt<12~\gevc$ (bottom-panels).
The $\Ntrk \in$ [1,8], [14,19] and [31,49] intervals are shown in the left, middle and right panels respectively. 
The fits to the candidate invariant mass distributions are also shown. 
\label{fig:DInvMass}
}
\end{center}
\end{figure}

The D-meson raw yields were extracted in each $\Ntrk$ and $\pt$ interval by means of a fit to the candidate invariant mass distributions (mass difference $\Delta M = M({\rm K} \pi\pi) -M({\rm K} \pi)$ for $\Dstar$). Similarly, the multiplicity-integrated raw yields were also evaluated for each $\pt$ interval. 
The $\Dzero$ and $\Dplus$ candidate invariant mass distributions were fitted with a function composed of a Gaussian for the signal and an exponential term that describes the background shape.
The $\Delta M$ distribution of $\Dstar$ candidates, which features a narrow peak at $\Delta M \simeq 145.4~\mev/c^2$~\cite{Agashe:2014kda}, was fitted with a Gaussian function for the signal and a threshold function multiplied by an exponential to model the background $\left(\sqrt{ \Delta M -M_{\pi} } \cdot e^{b (\Delta M -M_{\pi})}\right)$. 
The centroids of the Gaussians were found to be compatible with the world-average masses of the D mesons~\cite{Agashe:2014kda} in all multiplicity and $\pt$ intervals.
The widths of the Gaussian functions are independent of multiplicity and increase with increasing D-meson $\pt$, ranging between 10 and 20 $\mev/c^2$ for $\Dzero$ and $\Dplus$ and between 600 and 900 $\kev/c^2$ for $\Dstar$ mesons, consistent with the values obtained in simulations. In order to reduce the influence of statistical fluctuations, the raw yields were determined by constraining the D-meson line shape, its mass to the world-average D-meson mass, and its width to the value obtained from a fit to the invariant mass distribution in the
multiplicity-integrated sample, where the signal statistical significance is larger.  
Figure~\ref{fig:DInvMass} shows the $\Dzero$ and $\Dplus$ candidate invariant mass distribution, and $\Dstar$ mass difference distributions, for selected $\pt$ and multiplicity intervals. 
The extraction of the raw signal yields (sum of particle and antiparticle) was possible in five $\pt$ intervals from $1~\gev/c$ to $20~\gev/c$ for the $\Ntrk$ ranges reported in Table~\ref{tab:NchCorr}. 
The analysis covering the range $\Ntrk \in [1,49]$ exploited the minimum-bias triggered sample and was possible for the three D-meson species in three $\pt$ intervals in the range between $2$ and $12~\gev/c$. 
In addition, the $\Dzero$ signal was extracted in $\Ntrk \in [1,49]$ for $1<\pt<2~\gev/c$, and the $\Dstar$ signal was determined in three multiplicity intervals for $12<\pt<20~\gev/c$. 
The highest multiplicity interval $[50,80]$ was studied with the high multiplicity triggered sample via $\Dzero$ mesons for $2<\pt<4~\gev/c$ and the three D-meson species for $4<\pt<8~\gev/c$. 
The raw yield extraction in the remaining $\pt$ and multiplicity intervals for the different D-meson species was not performed due to the limited statistics in the analysed data sample and/or the large background.

\label{sec:Dreco}

\label{sec:Dsignal}

\subsection{Corrections}
\label{sec:Dcorrections}

The yields of D mesons were evaluated for each multiplicity and $\pt$ interval starting from the raw 
counts, $N_{\rm raw}$, which were divided by the reconstruction, topological and PID selection 
efficiencies for prompt D mesons, $\varepsilon_{\rm prompt \ D}$, and by the number of events analysed in the considered 
multiplicity interval, $N^{j}_{\rm event}$. 
The results are reported as the ratio of yields in each multiplicity interval, $(\dNDzdydpt)^j$, to the multiplicity-integrated (average) yield, $\langle \dNDzdydpt \rangle$, 
\begin{equation}
\left( \frac{ \dNDzdydpt } { \langle \dNDzdydpt \rangle } \right)^{j}= 
	\left(  \frac{1}{N^{j}_{\rm event}} \frac{N_{\rm raw \ \Dzero}^{j}}{\varepsilon^{j}_{\rm prompt \ \Dzero}} \right)
	 \Bigg/
	 \left( \frac{1}{N_{\rm MB \, trigger} \ \varepsilon_{\rm trigger}} \frac{ \langle N_{\rm raw \ \Dzero} \rangle }{ \langle \varepsilon_{\rm prompt \ \Dzero} \rangle} \right),
\label{eq:DCorrYields}
\end{equation}
where the index $j$ identifies the multiplicity interval.
The acceptance correction, defined as the fraction of D mesons within a given rapidity and $\pt$ interval whose decay particles are within the detector coverage, cancels in this ratio. 
D-meson raw yields have two components: the prompt D-meson contribution, and the feed-down contribution originating from B hadron decays. Equation~\ref{eq:DCorrYields} evaluates the yields of prompt D mesons under the assumption that the relative contribution to the D-meson raw yield due to the feed-down from B hadron decays does not depend on the multiplicity of the event, and is therefore cancelling in the ratio to the multiplicity-integrated values. 
This assumption is justified by the measurement of the multiplicity dependence of the B-hadron yields, via the non-prompt $\Jpsi$ fraction, presented in Sec.~\ref{sec:Jpsi} and by PYTHIA simulations. 
To evaluate the yields per inelastic collisions, the number of events used for the normalisation of the multiplicity-integrated yield has to be corrected for the fraction of inelastic collisions that are not selected by the minimum-bias trigger $N_{\rm MB \ trigger} / \varepsilon_{\rm trigger}$, with $\varepsilon_{\rm trigger} = 0.85^{+6\%}_{-3\%}$~\cite{Abelev:2012sea}. 
The results are also reported in Tables~\ref{tab:DAverageNoSigmaInel} 
and~\ref{tab:DvsNvzeroNoSigmaInel} without this trigger efficiency correction.
It was verified with PYTHIA 6.4.21~\cite{Sjostrand:2006za} Monte Carlo simulations that this minimum-bias trigger is 100\% efficient for D mesons in the kinematic range of this measurement. 

The D-meson efficiency corrections were determined with Monte Carlo simulations using the PYTHIA 6.4.21 event generator~\cite{Sjostrand:2006za} with Perugia-0 tune~\cite{Skands:2010ak}, and the GEANT3 transport code~\cite{Geant3}. The detector configuration and the LHC beam conditions were included, taking into account their evolution with time during the data taking period.
The $\varepsilon^{j}_{\rm prompt \ D}$ depends on the D-meson species and on $\pt$.
For prompt $\Dzero$ mesons it is 3--4\% in the $2<\pt<4~\gev/c$ interval and it increases up to 25--35\% for $\pt>8~\gev/c$, because less stringent topological selections were used at high $\pt$, where the combinatorial background is smaller.
The efficiency for feed-down D mesons is larger by about 20--30\% than for prompt D mesons. 
This is due to the fact that feed-down D mesons decay further away from the interaction vertex and are therefore more efficiently selected by the topological requirements.
The D-meson selection efficiency depends also on the multiplicity of charged particles produced in the collision, because the resolution on the position of primary vertex improves with increasing multiplicity, providing a better resolution of the variables used for the topological selections. 
For example, the $\Dzero$ selection efficiency in $2<\pt<4~\gev/c$ increases by about 40\% from the lowest to the highest multiplicity intervals considered in this analysis.

\subsection{Systematic uncertainties}
\label{sec:DSyst}

Several sources of systematic uncertainty that could affect the relative yields as expressed in Eq.~\ref{eq:DCorrYields} were studied. 
Only the raw yield extraction and the feed-down subtraction contribution were found to have an influence on the relative yields. 
The influence of the raw signal extraction from the invariant mass distribution was evaluated by using the raw yields obtained with different approaches to separate the signal from the combinatorial background. 
The contribution to the $\Dzero$ line shape of mis-identified K and $\pi$ pairs from $\Dzero$ decays, e.g.\ a $\Dzero \to {\rm K}^- \, \pi^+$ that passes the selection criteria as $\Dzerobar \to \pi^- \, {\rm K}^+$, was assumed to be the same in all multiplicity intervals and was neglected in this analysis.
Different background fit functions were considered (exponential, polynomial, linear for $\Dzero$ and $\Dplus$; threshold, $(\Delta M -M_{\pi})^b $, for $\Dstar$); the centroid and width of the Gaussians were left as free parameters in the fit instead of keeping them fixed to the values obtained from the multiplicity-integrated distribution; the raw yield was also extracted by counting the invariant mass histogram entries in a $\pm 3 \sigma$ interval around the peak after subtracting the background evaluated by fitting the distribution side bands (i.e.\ excluding the $\pm 3 \sigma$ interval around the centroid). 
The uncertainty was estimated from the stability of the ratio of the raw yields $N_{\rm raw \ \Dzero}^{j}/ \langle N_{\rm raw \ \Dzero} \rangle $, where the same raw yield extraction method was used in the multiplicity interval $j$ and for the multiplicity-integrated result. The assigned systematic uncertainty varies from 3\% to 15\% depending on the meson species, $\pt$ and multiplicity interval.

The efficiency corrections were calculated independently for each multiplicity interval. The multiplicity distribution of primary charged particles in the Monte Carlo simulation, $P(N_{\rm ch})$, was tuned to reproduce the measured charged-particle multiplicity~\cite{Aamodt:2010pp}. 
The efficiencies obtained with different Monte Carlo setups, that generate different initial multiplicity distributions, showed a good agreement in all multiplicity intervals. 
This effect was not considered as a source of systematic uncertainty. 

The D-meson decay tracks can be included or not in {\it (i)} the counting of the number of tracklets, resulting in a shift of the estimated multiplicity, and in {\it (ii)} the determination of the primary vertex position, which leads to a different resolution on the vertex position and of the geometrical variables used for the D-meson selection.
In the default configuration, the analysis was done excluding the D-meson decay tracks from the primary vertex determination and without excluding them from the multiplicity estimation.
To check for possible systematic effects due to the multiplicity determination, the analysis was repeated excluding the D-meson decay tracks from the multiplicity estimation, obtaining compatible results. 
Furthermore, the relative yields were determined without excluding the D-meson decay tracks from the primary vertex determination.
The influence of such variation is properly reproduced by Monte Carlo simulations, leading to a null effect on the corrected relative yields. Therefore this effect was not considered as a source of systematic uncertainty. 

The analysis was repeated for all D-meson species with different sets of topological selection criteria.
It was verified that the corrected relative D-meson yields as defined in Eq.~\ref{eq:DCorrYields} are not sensitive to this variation. 
This confirmed that the systematic uncertainty related to the topological selection description in the Monte Carlo cancels in the ratio. 
The influence of the PID strategy, which is based on the information of TPC and TOF detectors, was studied by also extracting the D-meson raw yields without PID selection criteria, which could be done only for D-meson $\pt>2~\gev/c$. 
The ratios of the relative raw yields, $N_{\rm raw \ \Dzero}^{j}/ \langle N_{\rm raw \ \Dzero} \rangle $, with and without PID selections were found to be compatible with unity.  
As a consequence, this effect was not considered as a source of systematic uncertainty.

As mentioned above, Eq.~\ref{eq:DCorrYields} describes the prompt corrected yields under the assumption that the fraction of prompt D mesons, $f_{\rm prompt}$, does not vary with the event multiplicity.
To estimate the uncertainty related to this assumption, the multiplicity integrated $f_{\rm prompt}$ factor was evaluated with the FONLL B-hadron production cross sections~\cite{Cacciari:2012ny}, the B~$\rightarrow$~D+$X$ decay kinematics from EvtGen~\cite{Lange2001152}, and the acceptance, reconstruction and selection efficiency of D mesons from B decays as described in~\cite{ALICE:2011aa}. 
The resulting $f_{\rm prompt}$ values are about 85--95\% depending on the D-meson $\pt$ and the applied selection criteria. 
The uncertainty due to the B feed-down contribution to the relative yields, $f_{\rm B} = 1-f_{\rm prompt}$, was evaluated assuming a linear increase of the fraction $f_{\rm B}^{j}/\langle f_{\rm B} \rangle$ from $1/2$ to $2$ from the lowest to the highest multiplicity interval. 
The resulting uncertainties vary with the $\pt$ and multiplicity range and are different for the three mesons. Typical values for intermediate $\pt$ at low multiplicity are $^{+5}_{-0}\%$, and at high multiplicity $^{+0}_{-20}\%$.

\subsection{Results}
\label{sec:Dresults}

\begin{figure}[!htb]
\begin{center}
\subfigure[D meson with $1<\pt<2~\gev/c$]{
	\label{fig:DYieldCorr_1_2}
	\includegraphics[width=0.475\columnwidth]{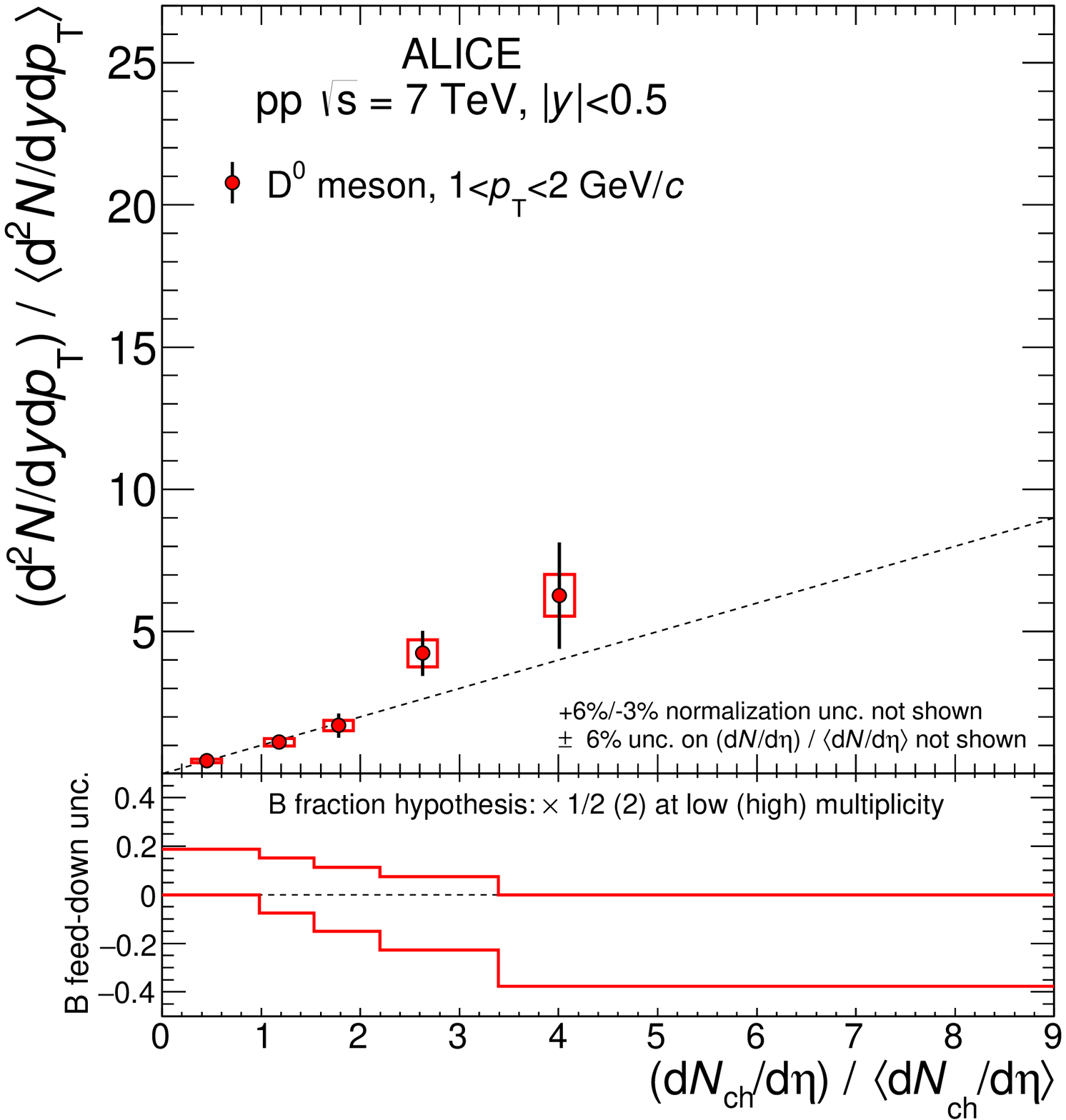}
	} 
\subfigure[D meson with $2<\pt<4~\gev/c$]{
	\label{fig:DYieldCorr_2_4}
	\includegraphics[width=0.475\columnwidth]{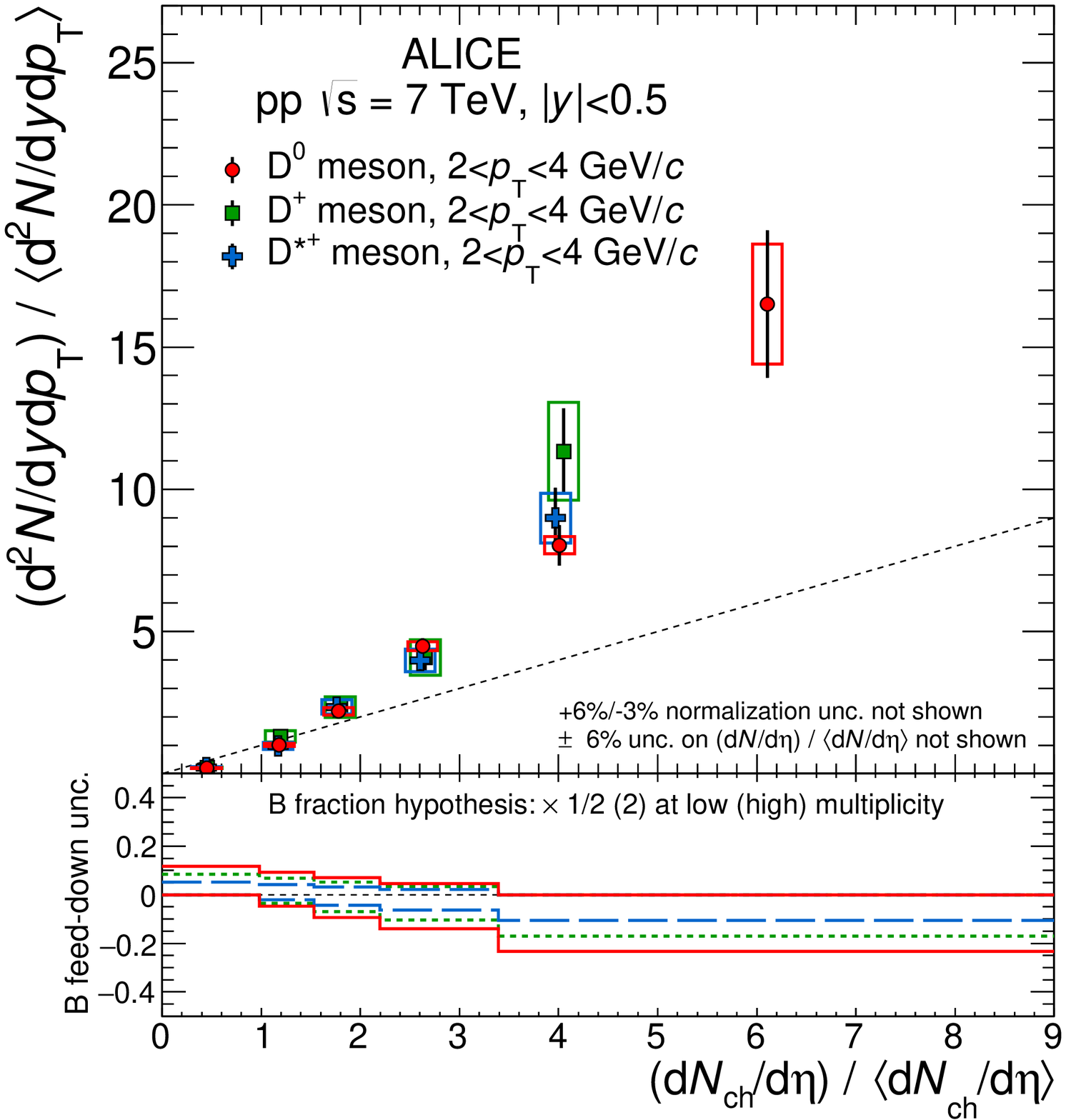}
	} 
\caption{$\Dzero$, $\Dplus$ and $\Dstar$ meson relative yields for each $\pt$ interval as a function of charged-particle multiplicity at central rapidity. 
The relative yields are presented on the top panels with their statistical (vertical bars) and systematic (boxes) uncertainties, except for the feed-down fraction uncertainty that is drawn separately in the bottom panels.
$\Dzero$ mesons are represented by red circles, $\Dplus$ by green squares, and $\Dstar$ by blue triangles. 
The position of the points on the abscissa is the average value of $(\dNdEta) \big/ \langle \dNdEta \rangle$. 
For $\Dplus$ and $\Dstar$ mesons the points are shifted horizontally by $1.5\%$ to improve the visibility. 
The diagonal (dashed) line is also shown to guide the eye.
\label{fig:DYieldCorr1}
}
\end{center}
\end{figure}
\begin{figure}[!htb]
\begin{center}
\subfigure[D meson with $4<\pt<8~\gev/c$]{
	\label{fig:DYieldCorr_4_8}
	\includegraphics[width=0.475\columnwidth]{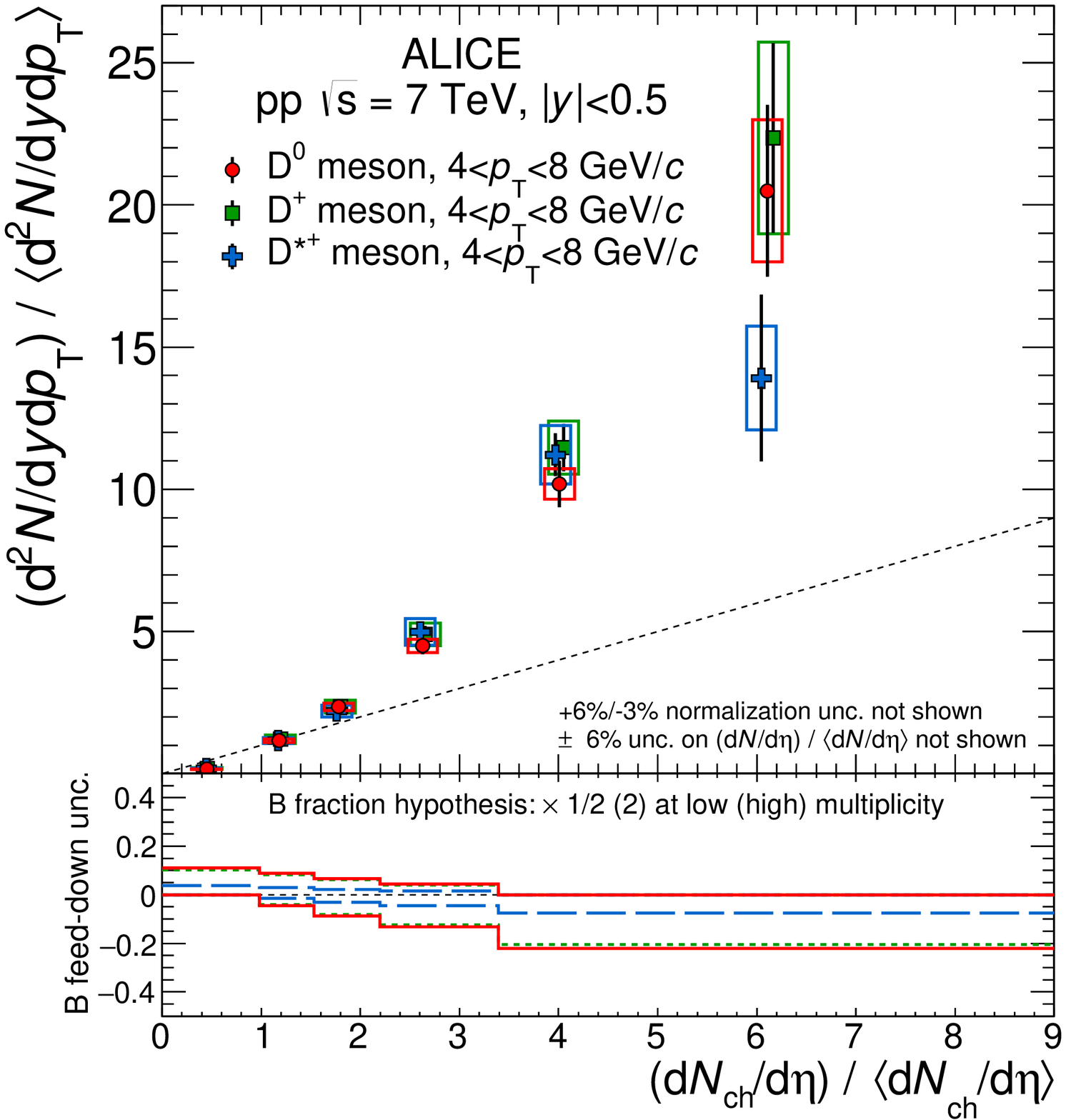}
	} 
\subfigure[D meson with $8<\pt<12~\gev/c$]{
	\label{fig:DYieldCorr_8_12}
	\includegraphics[width=0.475\columnwidth]{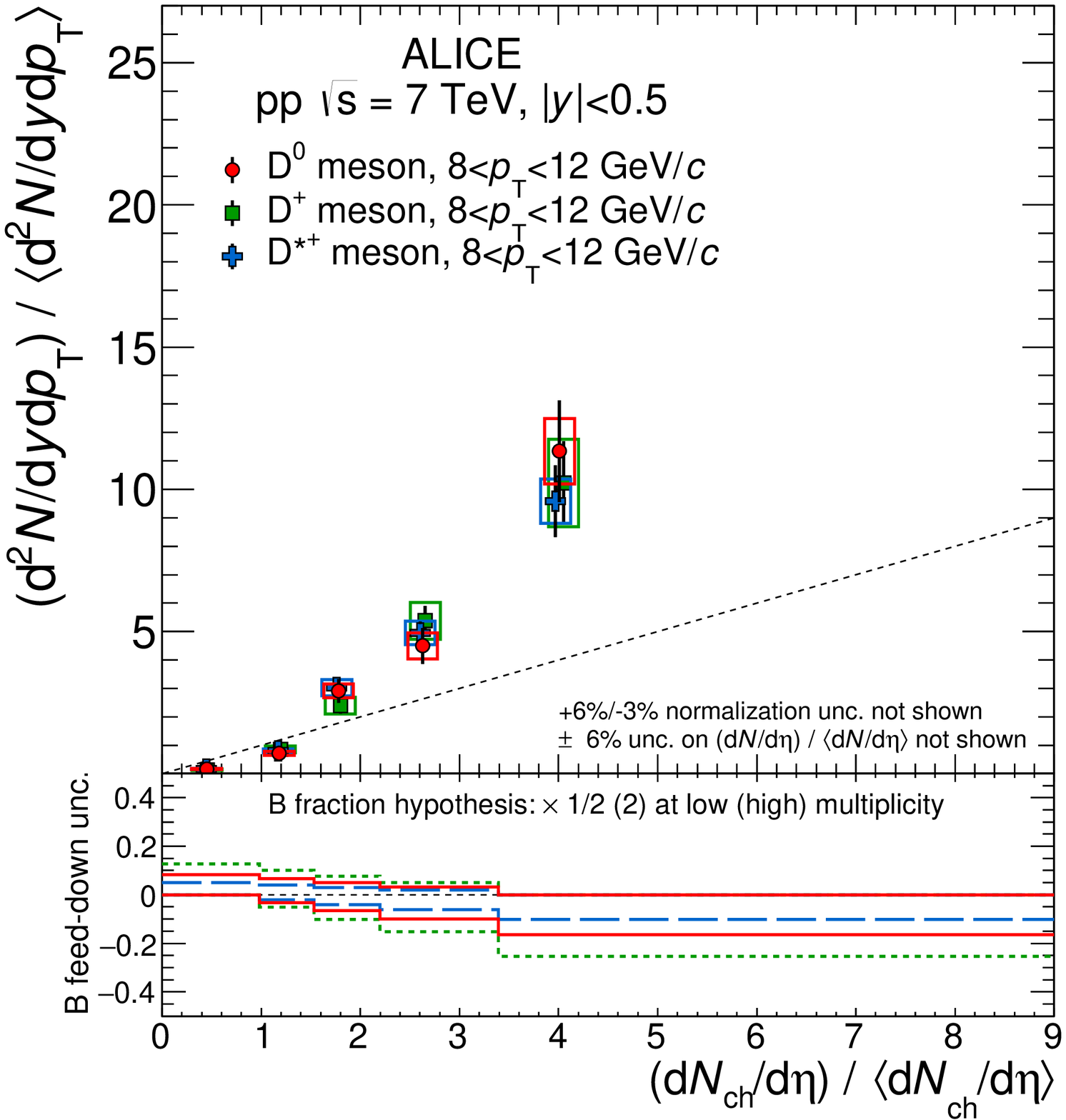}
	} 
\subfigure[D meson with $12<\pt<20~\gev/c$]{
	\label{fig:DYieldCorr_12_20}
	\includegraphics[width=0.475\columnwidth]{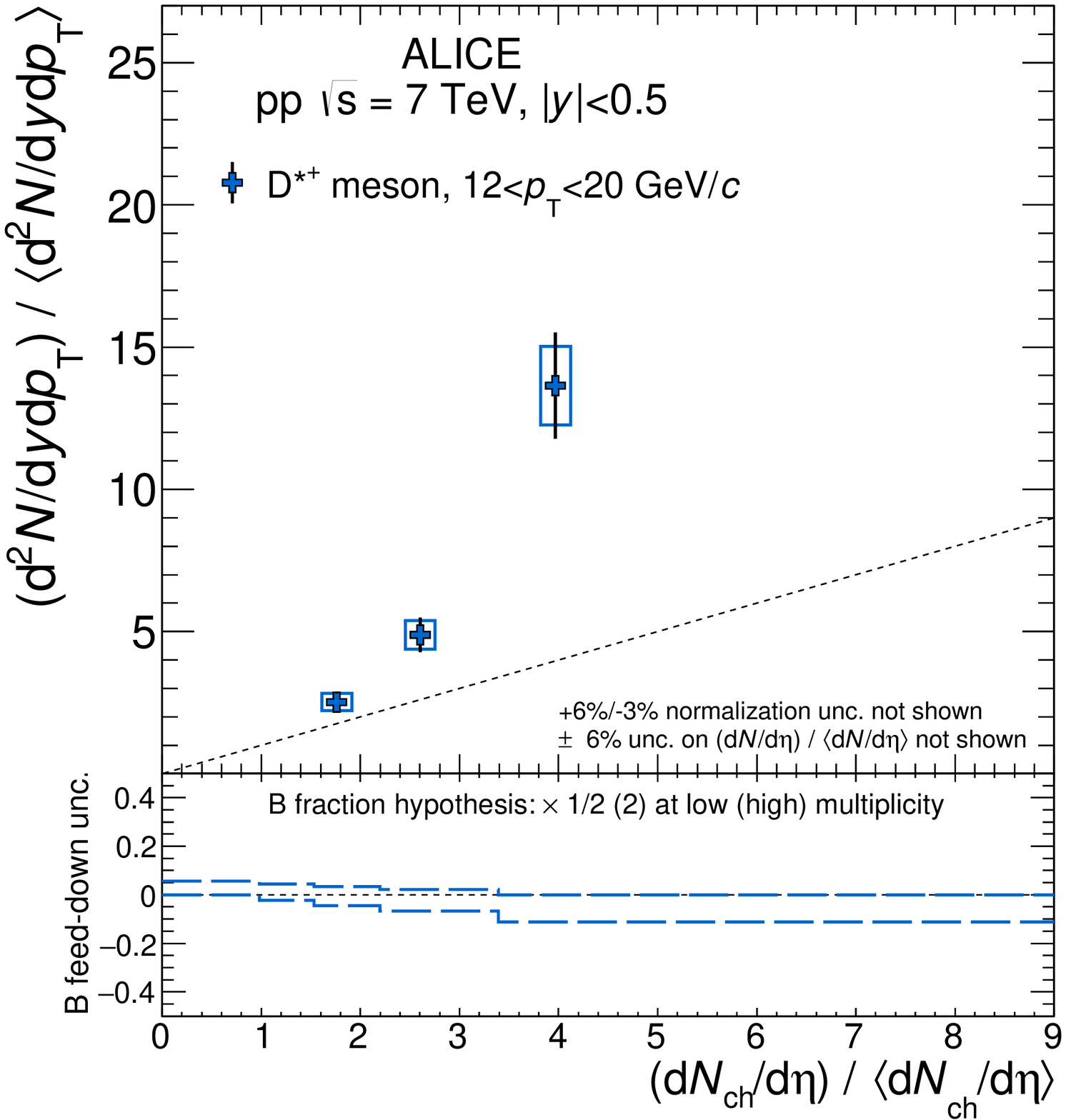}
	} 
\caption{$\Dzero$, $\Dplus$ and $\Dstar$ meson relative yields for each $\pt$ interval as a function of charged-particle multiplicity at central rapidity. 
The relative yields are presented on the top panels with their statistical (vertical bars) and systematic (boxes) uncertainties, except for the feed-down fraction uncertainty that is drawn separately in the bottom panels.
$\Dzero$ mesons are represented by red circles, $\Dplus$ by green squares, and $\Dstar$ by blue triangles. 
The position of the points on the abscissa is the average value of $(\dNdEta) \big/ \langle \dNdEta \rangle$. 
For $\Dplus$ and $\Dstar$ mesons the points are shifted horizontally by $1.5\%$ to improve the visibility. 
The diagonal (dashed) line is also shown to guide the eye.
\label{fig:DYieldCorr2}
}
\end{center}
\end{figure}

The results of the $\Dzero$, $\Dplus$ and $\Dstar$ meson relative yields for each $\pt$ interval are presented in Figs.~\ref{fig:DYieldCorr1} and~\ref{fig:DYieldCorr2} as a function of the relative charged-particle multiplicity. The relative yields are presented in the top panels with their statistical (vertical bars) and systematic (boxes) uncertainties except the uncertainty on the feed-down fraction, which is drawn separately in the bottom panels in the form of relative uncertainties. 
The position of the points on the abscissa is the average value of the relative charged-particle multiplicity, $(\dNdEta ) \big/ \langle \dNdEta \rangle$, for every $\Ntrk$ interval. 
The $\Dzero$, $\Dplus$ and $\Dstar$ meson relative yields are compatible in all $\pt$ intervals within uncertainties. 

\begin{figure}[!htb]
\begin{center}
\subfigure[$\pt$ dependence.]{
	\label{fig:DYieldCorrAv_pt}
	\includegraphics[width=0.475\columnwidth]{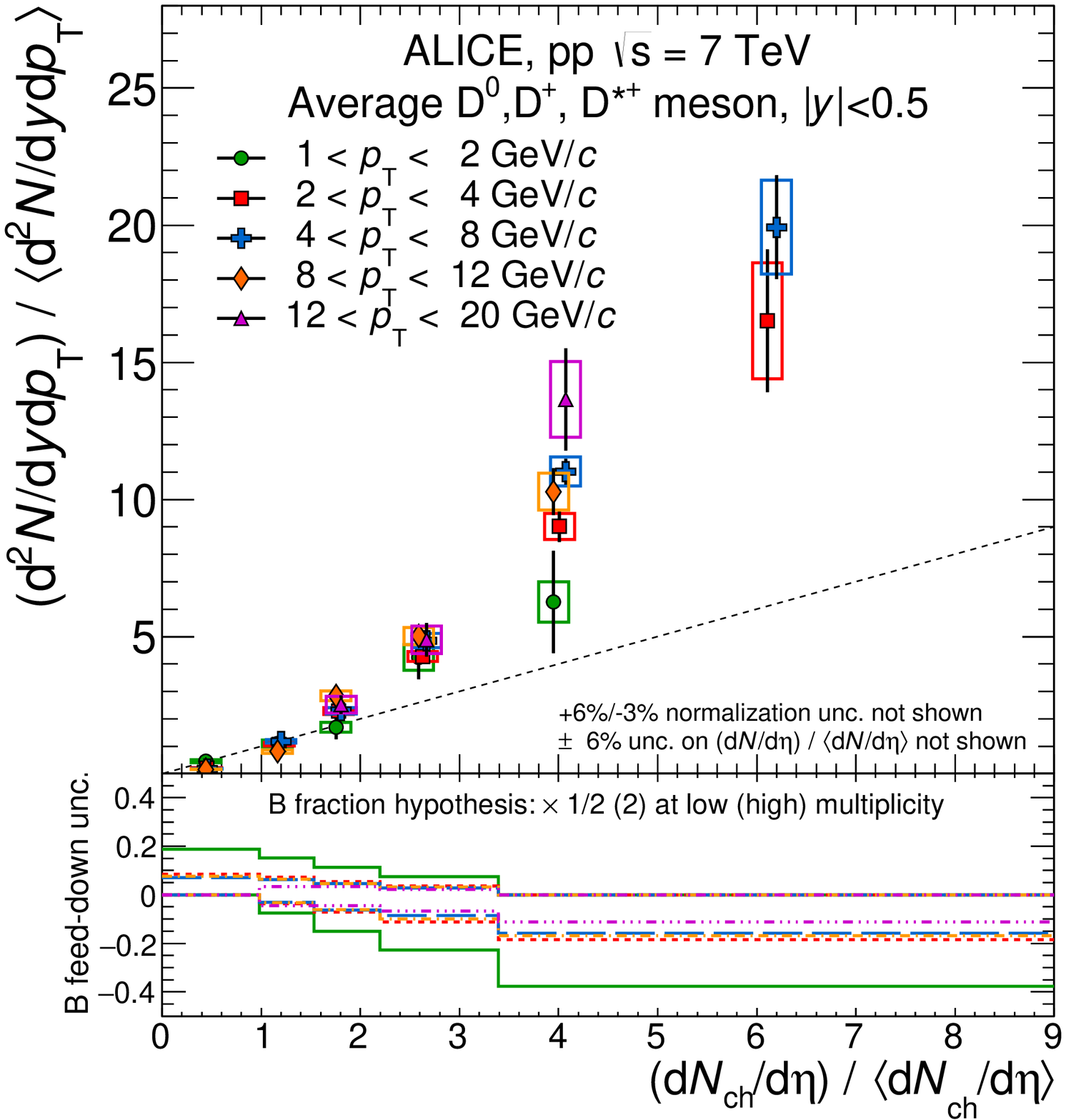}
	} 
\subfigure[Ratios of $\pt$ intervals vs the $2<\pt<4~\gev/c$.]{
	\label{fig:DYieldCorrAv_pt_ratio}
	\includegraphics[width=0.475\columnwidth]{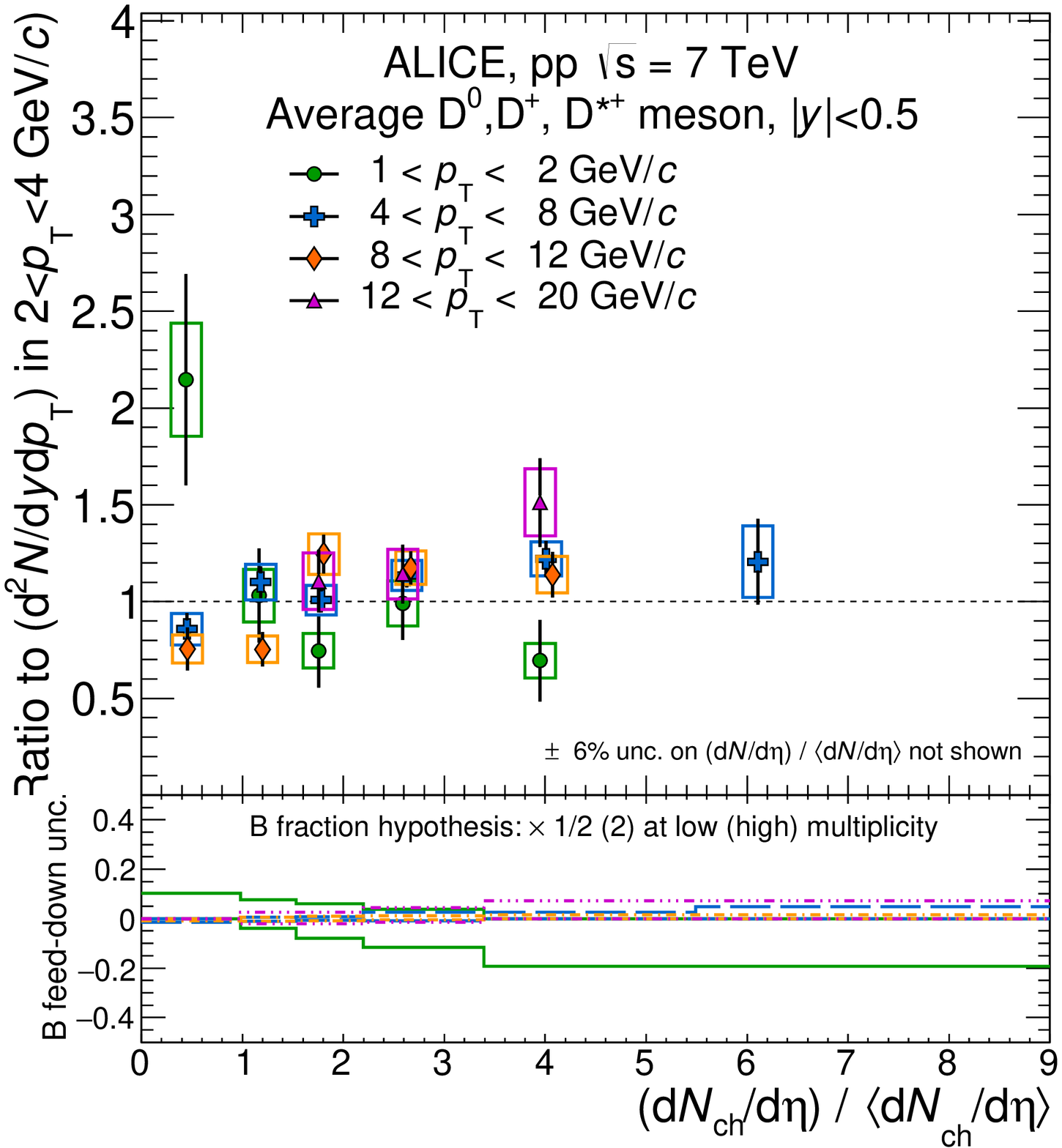}
	} 
\caption{Average of $\Dzero$, $\Dplus$ and $\Dstar$ relative yields as a function of the relative charged-particle multiplicity at central rapidity.
	(a)~Average of D-meson relative yields in $\pt$ intervals. 
	(b)~Ratio of the average relative yields in all $\pt$ intervals with respect to that of the $2<\pt<4~\gev/c$ interval.
	The results are presented in the top panels with their statistical (vertical bars) and systematic (boxes) uncertainties, except for the feed-down fraction uncertainty that is drawn separately in the bottom panels.
	The position of the points on the abscissa is the average value of $(\dNdEta) \big/ \langle \dNdEta \rangle$. 
	For some $\pt$ intervals the points are shifted horizontally by $1.5\%$ to improve the visibility. 
	The dashed lines are also shown to guide the eye, a diagonal on (a) and a constant on (b).
\label{fig:DYieldCorrAverage}
}
\end{center}
\end{figure}

The average of $\Dzero$, $\Dplus$ and $\Dstar$ relative yields was computed for each $\pt$ interval using as weights the inverse square of their relative statistical uncertainties. 
The yield extraction uncertainty was considered as uncorrelated systematic uncertainty. The feed-down fraction systematic uncertainty was treated as a correlated systematic uncertainty. 
The average of the D-meson relative yields for all $\pt$ intervals is summarised in Tables~\ref{tab:DAverageNoSigmaInel} and~\ref{tab:DAverage}, and presented in Fig.~\ref{fig:DYieldCorrAv_pt}. The relative D-meson yields increase with the charged-particle multiplicity by about a factor of 15 in the range between 0.5 and six times $\langle \dNdEta \rangle$. 
Figure~\ref{fig:DYieldCorrAv_pt_ratio} shows the ratios of the average of the 
D-meson relative yields in various $\pt$ intervals with respect to the $2<\pt<4~\gev/c$ interval values. 
The yield enhancement is independent of transverse momentum within the uncertainties of the measurement.

\subsubsection{Studies with the charged-particle multiplicity at forward rapidity}
\label{sec:DresultsVzero}

\begin{figure}[!htbp]
\begin{center}
	\includegraphics[width=0.475\columnwidth]{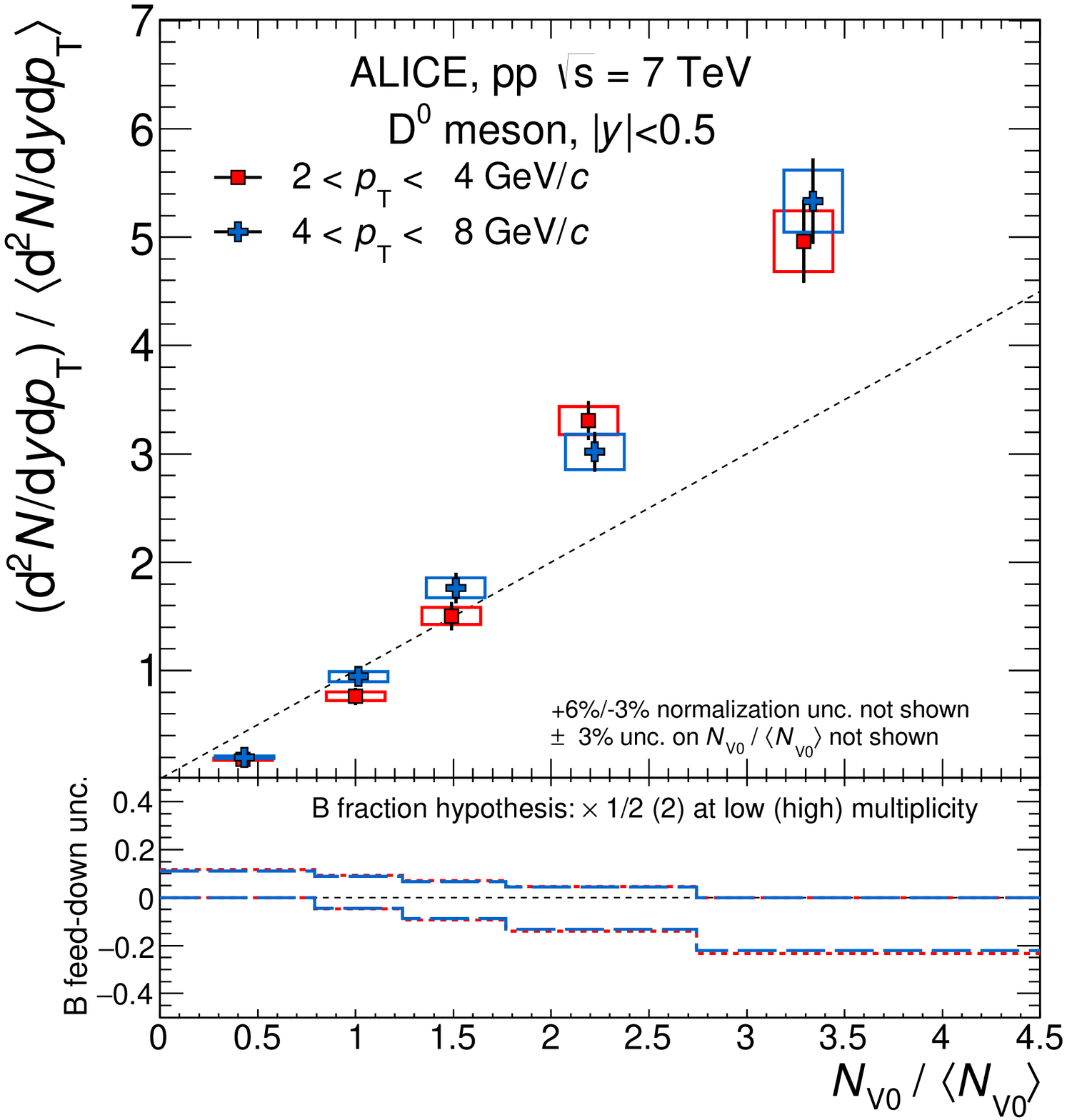}
\caption{$\Dzero$ meson relative yields at $|y|<0.5$ for two $\pt$ intervals as a function of the relative charged-particle multiplicity, $\Nvzero$, measured at $-3.7 < \eta < -1.7$ and $2.8 < \eta < 5.1$. 
The relative yields are presented on the top panels with their statistical (vertical bars) and systematic (boxes) uncertainties, except the uncertainty on the feed-down fraction which is drawn separately in the bottom panels.
The position of the points on the abscissa is at the average value of $\Nvzero \big/ \langle \Nvzero \rangle$,
shifted by $1.5\%$ to improve the visibility. 
The diagonal (dashed) line is also shown to guide the eye.
\label{fig:DzeroYieldCorrVzero}
}
\end{center}
\end{figure}

In the analysis described above, D-meson yields were measured in the same rapidity interval as the charged-particle multiplicity. 
This could lead to a bias if the particles produced in the charm-quark fragmentation and in the D-meson decay would amount to a large fraction of the measured charged particles. 
In order to study this possible bias, the measurement of the $\Dzero$ yields at central rapidity was also performed as a function of the relative charged-particle multiplicity at forward-rapidity.
The charge collected by the V0 scintillator counters, covering $-3.7 < \eta < -1.7$ and $2.8 < \eta < 5.1$, was used as multiplicity estimator in this case. 
The multiplicity value $\Nvzero$ was evaluated by dividing the collected charge by the expected average minimum-ionizing-particle charge. 
The $\Dzero$ yields were evaluated in intervals of $\Nvzero$, and corrected as previously described and summarised in Eq.~\ref{eq:DCorrYields}. The relative yields of $\Dzero$ mesons are presented in Fig.~\ref{fig:DzeroYieldCorrVzero} as a function of the  relative mean multiplicity measured with the V0 counters, $\Nvzero \big/ \langle \Nvzero \rangle$. 
The statistical (systematic) uncertainties are represented by the vertical bars (empty boxes). 
The systematic uncertainties due to the raw yield extraction and the B feed-down contribution were determined as explained in Sec.~\ref{sec:DSyst}.
The uncertainty due to the unknown feed-down fraction evolution with the charged-particle multiplicity is drawn separately in the bottom panels. 
The points are located on the $x$-axis at the average value of the relative mean multiplicity, $\Nvzero \big/ \langle \Nvzero \rangle$. The uncertainty on the mean multiplicity values, $\Nvzero$, was determined by comparing the mean and median values of the distributions. It was found to be below 3\% for each multiplicity interval, and about $24\%$ for the multiplicity integrated value. The uncertainty on $\Nvzero \big/ \langle \Nvzero \rangle$ is not displayed on this figure. 
These results are also summarised in Tables~\ref{tab:DvsNvzeroNoSigmaInel} and~\ref{tab:DvsNvzero}. 
The $\Dzero$ relative yields increase with the relative uncorrected multiplicity at forward rapidity, as measured with the V0 detector. 
The results in the $2<\pt<4~\gevc$ and $4<\pt<8~\gevc$ intervals are compatible within uncertainties. 
The results with the V0 multiplicity estimator indicate that the increase of the D-meson yield with the event multiplicity observed with the mid-rapidity estimator is not related to the fact that charmed mesons, originating from the fragmentation of charm quarks produced in hard partonic scattering processes, and the charged particle multiplicity are measured in the same pseudo-rapidity range. 
A qualitatively similar increasing trend of D-meson yield with multiplicity is indeed observed also when an $\eta$ gap is introduced between the regions where the D-mesons and the multiplicity are measured.

\section{Non-prompt $\Jpsi$ analysis}
\label{sec:Jpsi}

\subsection{Non-prompt $\Jpsi$ reconstruction}
\label{sec:JpsiAnalysis}

The fraction of non-prompt $\Jpsi$ in the inclusive $\Jpsi$ yields, $f_{\rm B}$, was measured as a function of the charged-particle multiplicity by studying displaced $\Jpsi$ mesons that decay into electron pairs in the rapidity range $|y|<0.9$. 
This measurement, combined with the inclusive $\Jpsi$ relative yield~\cite{Abelev:2012rz}, provides the multiplicity dependence of the production of beauty hadrons. 
$\Jpsi$ candidates were formed by combining pairs of opposite-sign electron tracks. 
The tracks were required to have $\pt>1~\gevc$, at least 70 (out of a maximum of 159) associated space points in the TPC with a $\chi^2/{\rm ndf}$  of the momentum fit lower than 2, 
%
 and to point back to the primary interaction vertex within $1$~cm in the transverse plane. The tracks were also required to have at least one associated hit in the SPD detector, with the constraint that one of the two tracks should have a hit in the first SPD layer. 
Electron identification was based only on the TPC information. A selection of $\pm3\sigma$ around the expected mean values of the specific energy deposit $\dEdx$ for electrons was used. To further reduce the background, a $\pm3.5\sigma$ ($\pm3\sigma$) exclusion band around the expected mean specific energy deposit for pions (protons) was also applied. 
In order to reduce the combinatorial background, electron candidates compatible, together with a positron candidate,
with being products of $\gamma$-conversions (invariant mass below $100~\mev/c^2$) were removed.

The measurement of $f_{\rm B}$ is based on a statistical discrimination of $\Jpsi$ mesons produced at a secondary vertex displaced from the primary \pp~collision vertex.
The signed projection of the $\Jpsi$ flight distance onto its transverse momentum vector, $\vec{\pt}$, was constructed as $L_{\rm xy} = \left( \vec{L} \cdot \vec{\pt} \right) / \pt$, where $\vec{L}$ is the vector from the primary vertex to the $\Jpsi$ decay vertex.
The pseudo-proper decay length $x = \left(c \cdot L_{\rm xy} \cdot m \right) \big/ \pt$ was calculated from the observed decay length using the world-average  $\Jpsi$ mass $m(\Jpsi)=3096.916 \pm 0.011~\mevcc$~\cite{Agashe:2014kda}. 
The fraction of non-prompt $\Jpsi$ can be determined from a 2-dimensional un-binned log-likelihood fit to $x$ and the unlike-sign di-electron invariant mass distributions. The fit procedure and the functions used to describe the invariant mass and the pseudo-proper decay length distributions were introduced in~\cite{Abelev:2012gx}.

\label{sec:dataAnalysis}
The fraction of non-prompt $\Jpsi$ as a function of the relative charged-particle multiplicity was determined for $\pt>1.3~\gevc$ in five multiplicity intervals in the $\Ntrk$ range $[4,49]$. 
The $\Ntrk \in [1,3]$ range was excluded from this analysis due to the poor pseudo-proper decay length resolution, $R(x)$, and the presence of a bias in the determination of $x$ in the case of non-prompt candidates. 
The resolution of the pseudo-proper decay length is determined with Monte Carlo simulations evaluating the RMS of the $x$ distributions of reconstructed promptly produced $\Jpsi$ mesons. 
The event primary vertex can be computed with or without removing the decay tracks of the $\Jpsi$ candidates. 
The removal of the decay tracks causes a degradation of the resolution on $x$, especially in the low-multiplicity intervals, as a consequence of the lower precision in the determination of the primary vertex with a reduced number of tracks. 
For simulated events with non-prompt $\Jpsi$, the removal of the decay tracks also results in a shift of the primary vertex position away from the secondary decay vertex of the beauty hadrons, which is reflected in a systematic shift of the mean of the $x$ distribution. 
However, one should consider that beauty quarks are always produced in pairs: the two decay tracks from the non-prompt-$\Jpsi$, when included, pull the primary vertex towards the beauty hadron decay vertex, but the charged tracks from the decay of the second beauty quark, which enter in the barrel acceptance, pull the primary vertex in the opposite direction. 
The shift is larger in the lowest multiplicity bin where it reaches about \mbox{35 $\mu$m}. 
This bias is reduced when the $\Jpsi$ decay tracks are kept in the evaluation of the primary vertex. 
\begin{figure}[tbh]
\begin{center}
\includegraphics[width=0.95\textwidth]{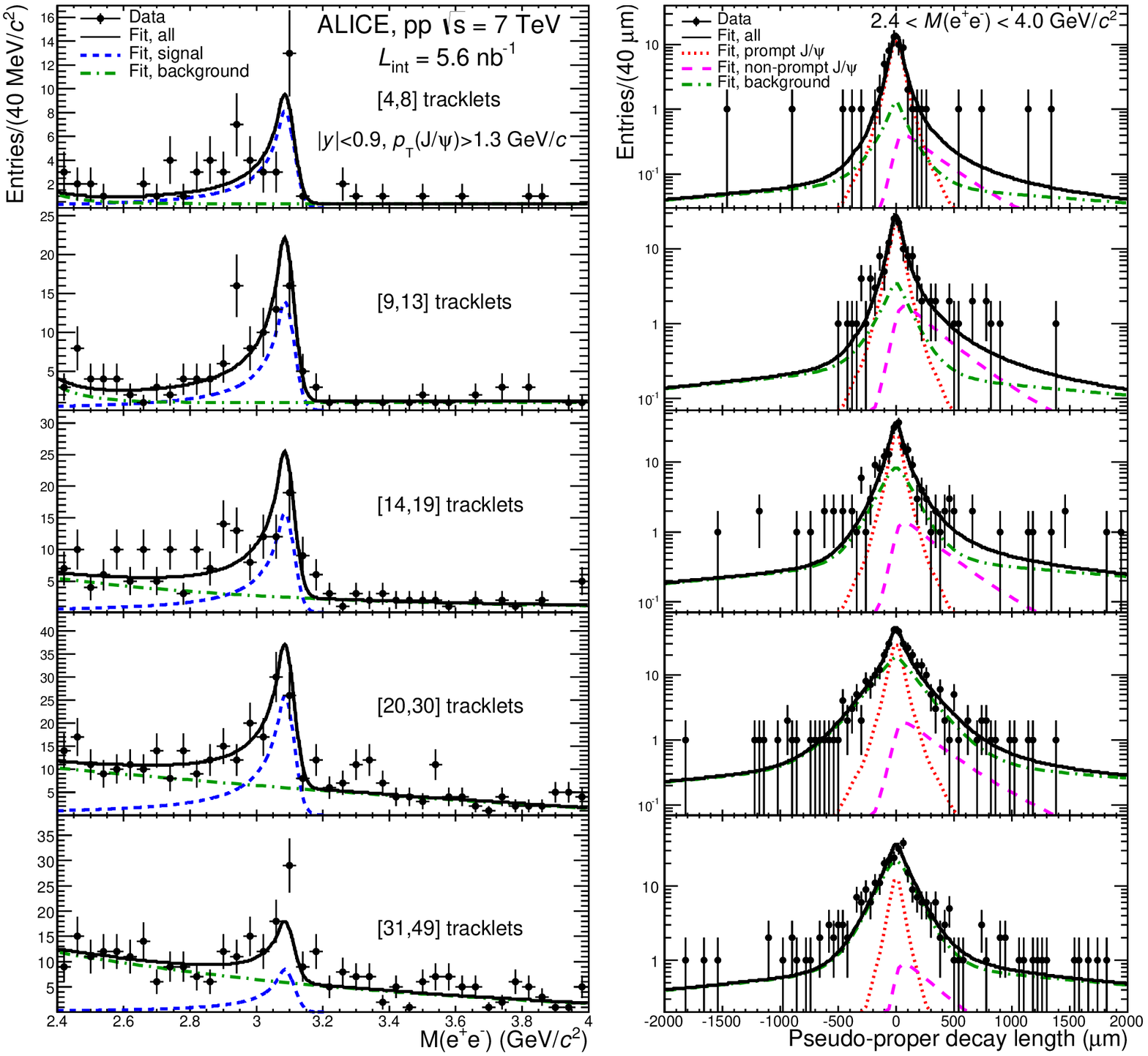}
\caption{\label{fig:LikelihoodFit}
$\Jpsi$ invariant mass and pseudo-proper decay length distributions in several multiplicity intervals with superimposed the likelihood fit results. The contributions of the signal, the background and their sum are represented with dashed, dot-dashed and full lines, respectively.
In addition, the pseudo-proper decay length figures include the prompt and non-prompt contributions to the inclusive yields with dotted and long-dashed lines. 
}
\end{center}
\end{figure}
The effect of the bias, estimated with Monte Carlo simulations, is a reduction of the measured 
$\fB$\footnote{
This shift would be greater than \mbox{50 $\mu$m} in the $\Ntrk$ interval $[1,3]$, leading to a large bias on the extracted $\fB$ value (up to 35\%). The correction for this bias would introduce a large systematic uncertainty.
} by about 20\% for events with $\Ntrk$ = 4, and it becomes negligible for $\Ntrk >10$. 
Therefore, the primary vertex was computed considering all reconstructed tracks. To correct for the remaining bias, a modification in the resolution function, $R(x)$, used to describe the non-prompt $\Jpsi$ in the likelihood fit function was introduced, which depends on $\Ntrk$. In particular, the shape of the resolution function was adjusted to obtain a good matching between the function used to describe the non-prompt $\Jpsi$ in the likelihood fit (a convolution of a template of the $x$ distribution of $\Jpsi$ from beauty hadron decays with the 
resolution function~\cite{Abelev:2012gx}) and
the pseudo-proper decay length distribution of reconstructed secondary $\Jpsi$ from Monte Carlo simulations. 

Figure~\ref{fig:LikelihoodFit} presents the invariant mass and pseudo-proper decay length distributions for $\pt>1.3~\gevc$ for each multiplicity interval together with a projection of the result of the log-likelihood fit.

\subsection{Corrections}

For all multiplicity intervals, the measured fraction of non-prompt $\Jpsi$, $\fB^{\prime}$, was corrected using the acceptance and reconstruction efficiency of prompt, $\langle Acc \times \varepsilon \rangle_{\rm prompt}$, and non-prompt $\Jpsi$, $\langle Acc \times \varepsilon \rangle_{\rm B}$, as
\begin{equation}
\label{CorrFb}
\fB= \left( 1 + \frac{1-\fB^{\prime}}{\fB^{\prime}} \cdot \frac{\langle Acc \times \varepsilon \rangle_{\rm B}}{\langle Acc \times \varepsilon \rangle_{\rm prompt}} \right)^{-1}.  
\end{equation} 
Here all terms refer to non-prompt $\Jpsi$ with $\pt>1.3~\gevc$. 
The corrections for acceptance and efficiency were computed using Monte Carlo simulations using the GEANT3 transport code~\cite{Geant3}. Prompt $\Jpsi$ were generated with a $\pt$ distribution extrapolated from CDF measurements~\cite{Acosta:2004yw} and a $y$ distribution parameterised with the Colour Evaporation Model (CEM)~\cite{fritzsch1977producing, Amundson:1996qr}. Beauty hadrons were generated using the PYTHIA 6.4.21 event generator~\cite{Sjostrand:2006za} with Perugia-0 tune~\cite{Skands:2009zm}. 
The acceptance times efficiency values for prompt and non-prompt $\Jpsi$ have a minimum of 8\% at $\pt = 2~\gevc$ and a broad maximum of 12\% at $\pt = 7~\gevc$~\cite{Aamodt:2011incl}. The relative difference in efficiency between prompt and non-prompt $\Jpsi$ is only about 3\%.
The ratio $\langle Acc \times \varepsilon \rangle_{\rm B} / \langle Acc \times \varepsilon \rangle_{\rm prompt}$ is assumed to be independent of multiplicity.
The uncertainty related to this assumption is discussed in the next section. 

The measured non-prompt $\Jpsi$ fractions were extrapolated from $\pt>1.3~\gevc$ down to $\pt=0$ using
\begin{equation}
\fB^{\rm extr}(\pt>0) = \alpha^{\rm extr}  \cdot \fB(\pt>1.3~\gevc) ; \qquad
\alpha^{\rm extr} = \frac{\fB^{\rm model}(\pt>0)}{\fB^{\rm model}(\pt>1.3~\gevc)}, 
\end{equation}
where $\fB^{\rm model}$ represents a functional form modelled on existing data. It was calculated as the ratio of the differential cross section of non-prompt $\Jpsi$, as obtained with FONLL calculations~\cite{Cacciari:2012ny}, to that of 
inclusive $\Jpsi$, parameterised by the phenomenological function defined in~\cite{Bossu:2011qe}:
\begin{equation}
\fB^{\rm model}(\pt) =  \frac{{\rm d}^2 \sigma^{\rm FONLL}_{\Jpsi \leftarrow h_{B}}}{{\rm d}y{\rm d}\pt} \, \Big/ \, \frac{{\rm d}^2 \sigma_{\Jpsi}^{\rm phenom}}{{\rm d}y{\rm d}\pt} . 
\end{equation}
A combined fit to the existing results of $\fB$ in pp collisions at 7 TeV~\cite{Abelev:2012gx,Aad:2011sp,Khachatryan:2010yr,Aaij:2011jh} in the rapidity bin closest to central rapidity was performed to determine the parameters of the phenomenological parameterisation. 
The extrapolation factor obtained is \mbox{$\alpha^{\rm extr} = 0.99^{+0.01}_{-0.03}$}. Its uncertainties were determined by repeating the fit by {\it (i)} excluding the LHCb data points at forward rapidities, and {\it (ii)} using for the non-prompt $\Jpsi$ cross section the upper and lower uncertainty bands of the FONLL predictions, obtained by varying the factorisation and renormalisation scales, instead of the central values. The uncertainties were determined by the maximum and minimum $\alpha^{\rm extr}$ values obtained from these fit variations. 
The $\fB$ fractions in all multiplicity intervals were extrapolated using the same $\alpha^{\rm extr}$ value, evaluated from the fit of the multiplicity integrated measurements.

\subsection{Systematic uncertainties}
\label{sec:JpsiSystematics}

The systematic uncertainty introduced by the experimental resolution on the primary vertex position was evaluated by repeating the fitting procedure in two alternative ways: 
{\it (i)} the primary vertex was evaluated without removing the decay tracks of the $\Jpsi$ candidates. The fit was performed using the standard resolution function for non-prompt $\Jpsi$, that does not depend on multiplicity, but the $x$ distribution of the non-prompt $\Jpsi$ was shifted by a multiplicity-dependent value, which was determined by the Monte Carlo simulation. 
{\it (ii)} The event primary vertex was computed after removing the decay tracks of the $\Jpsi$ candidates and the fit was performed using the corresponding degraded resolution function $R(x)$ and without any shift. 
The resulting uncertainties decrease with increasing multiplicity, ranging from 19\% in the lowest multiplicity interval to 3\% at the highest multiplicities.

The uncertainty related to the extrapolation of $\fB$ from $\pt>1.3~\gevc$ to $\pt>0$ was estimated with the method discussed above and it is about 3\%. This uncertainty was assumed to be uncorrelated among the multiplicity intervals.

The resolution function used in the fits is based on Monte Carlo simulations, which might introduce systematic effects. These were estimated by repeating the log-likelihood fits modifying the resolution function, $R(x)$, according to 
$\left( 1 / (1+\delta) \right) \cdot R \left( x / (1+\delta) \right)$,
where $\delta$ is the relative variation of the RMS of the resolution function,
and it was varied from $-0.1$ to $+0.1$ to take into account the uncertainties
in the Monte Carlo description. 
The systematic uncertainty due to the resolution function increases 
with multiplicity from $8\%$ to $20\%$. 

The $\pt$ distribution of the signal candidates (prompt and non-prompt $\Jpsi$) could depend on the event multiplicity which could affect the shape of the resolution function which depends on the $\Jpsi$ $\pt$. The average $\pt$ of the signal candidates was estimated from data in each multiplicity interval and found to be constant as a function of event multiplicity within statistical uncertainties about $\pm10\%$. 
The influence of a $\langle \pt \rangle$ variation on the resolution function was determined using Monte Carlo simulations: the $\pt$ distribution was changed, considering softer or harder $\pt$ distributions, in order to obtain a $\pm10\%$ variation of the $\langle \pt \rangle$. 
The corresponding variations obtained for the RMS of the resolution function are $+7\%$ and $-8.5\%$ for the softer and harder $\pt$ distribution, respectively. 
The latter variations are within those quoted for the resolution function ($\pm$10\%), therefore no additional uncertainty was included.

The acceptance times efficiency values of prompt and non-prompt $\Jpsi$  reconstructed for $\pt>1.3~\gevc$ are of the order of 10\% and differ by 3\%.   
The influence of the $\pt$ shape assumed in the simulation on the ratio $\langle Acc \times \varepsilon \rangle_{\rm B} / \langle Acc \times \varepsilon \rangle_{\rm prompt}$ was 
evaluated by varying the average $\pt$ of the simulated $\Jpsi$ distributions within $\pm$50\%. 
A $1\%$ variation in the acceptance was obtained both for prompt and non-prompt $\Jpsi$. The corresponding variation obtained on $\fB$ 
through the Eq.~\ref{CorrFb} is about 1\%.

The pseudo-proper decay length shape of the combinatorial background was determined by a fit to the $x$ distribution of the candidates in the sidebands of the invariant mass~\cite{Abelev:2012gx}. By varying the fit parameters within their errors an envelope of distributions was obtained, whose extremes were used in the likelihood fit to estimate the systematic uncertainty. It increases slightly with multiplicity, ranging from $1\%$ to $5\%$. 

The uncertainty on the background invariant mass shape, which was determined by fits to the invariant mass distributions of opposite-sign candidates in each multiplicity bin, was evaluated by using like-sign distributions instead, adopting the same procedure as described in~\cite{Abelev:2012gx}. The systematic uncertainty is about $7\%$, independent of the charged-particle multiplicity. 

The shape of the $x$ distribution of $\Jpsi$ from b-hadrons was evaluated using PYTHIA 6.4.21~\cite{Sjostrand:2006za}. 
The systematic uncertainty on its shape was computed by {\it (i)} changing the b-hadron decay kinematic, using EvtGen~\cite{Lange2001152} instead of PYTHA 6.4.21 or {\it (ii)} by assuming a harder and a softer b-hadron $\pt$ distribution, resulting in a $\langle \pt \rangle$ variation of about $\pm15\%$. 
The resulting systematic uncertainty is about $3\%$, constant with multiplicity. 

The signal invariant mass shape was fixed from the Monte Carlo simulation which includes the detector resolution effects and the radiative decays using the EvtGen~\cite{Lange2001152} package. The effect on the invariant mass signal shape due to the uncertainty on the detector material was studied with dedicated Monte Carlo simulations, where the detector material budget was varied by $\pm 6\%$ with respect to the nominal values~\cite{Abelev:2014ypa,Abelev:2012cn}. 
The resulting systematic uncertainty on $\fB$ is $3\%$ in the lowest event multiplicity interval and $5\%$ in the highest one. 

The systematic uncertainties on the pseudo-proper decay length of the combinatorial background, on the $\pt$-extrapolation uncertainty $\alpha^{\rm extr}$ and on the invariant mass shape of background are found or, in the case of $\alpha^{\rm extr}$, assumed to be uncorrelated among multiplicity intervals. The remaining systematic uncertainties are (fully or partially) correlated in different multiplicity intervals.

\subsection{Results}
\label{sec:Jpsiresults}

The relative yield of $\Jpsi$ from beauty hadron decays as a function of the charged-particle multiplicity was evaluated from the inclusive $\Jpsi$ yield and the fraction of non-prompt $\Jpsi$ per multiplicity interval: 
\begin{equation}   
\frac{ {\rm d} N_{\Jpsi}^{\rm non-prompt} / {\rm d}y }{ \left\langle  {\rm d} N_{\Jpsi}^{\rm non-prompt} / {\rm d}y \right\rangle }  =
\frac{ {\rm d} N_{\Jpsi} / {\rm d}y }{ \left\langle  {\rm d} N_{\Jpsi} / {\rm d}y \right\rangle }  \cdot
 \frac{\fB }{\left\langle \fB \right\rangle}.
\label{eq:JpsiYields}
\end{equation}
$\fB$ is the fraction of non-prompt $\Jpsi$ in each multiplicity interval,
$\langle \fB \rangle$ is the fraction in the multiplicity integrated sample~\cite{Abelev:2012gx}, 
and $({\rm d}N_{\Jpsi}/{\rm d}y) \big/ \langle {\rm d}N_{\Jpsi}/{\rm d}y \rangle$ is the inclusive $\Jpsi$ relative yield measured for \mbox{$\pt>0$} in each multiplicity interval normalized to its value in inelastic \pp~collisions~\cite{Abelev:2012rz}. 
All these quantities were measured using the same data sample and the statistical correlations were taken into account. 
In the first charged-particle multiplicity class $\Ntrk \in [4,8]$, which is used for the non-prompt $\Jpsi$ analysis presented here, the relative yield of inclusive $\Jpsi$ normalized to the inelastic cross section is $({\rm d}N_{\Jpsi}/{\rm d}y) \big/ \langle {\rm d}N_{\Jpsi}/{\rm d}y \rangle = 0.41  \pm 0.07\,  {\rm (stat)} \pm 0.01 \, {\rm (syst)}$.
The values of $\fB$ extrapolated to $\pt >0$ were used in Eq.~\ref{eq:JpsiYields}, providing the non-prompt $\Jpsi$ relative yields for $\pt >0$.  
The relative yields of inclusive $\Jpsi$ were also recomputed for $\pt >1.3~\gevc$ and no difference was observed with respect to those for $\pt >0$ within the uncertainties.  

\begin{figure}[!htb]
\begin{center}
	\includegraphics[width=0.475\columnwidth]{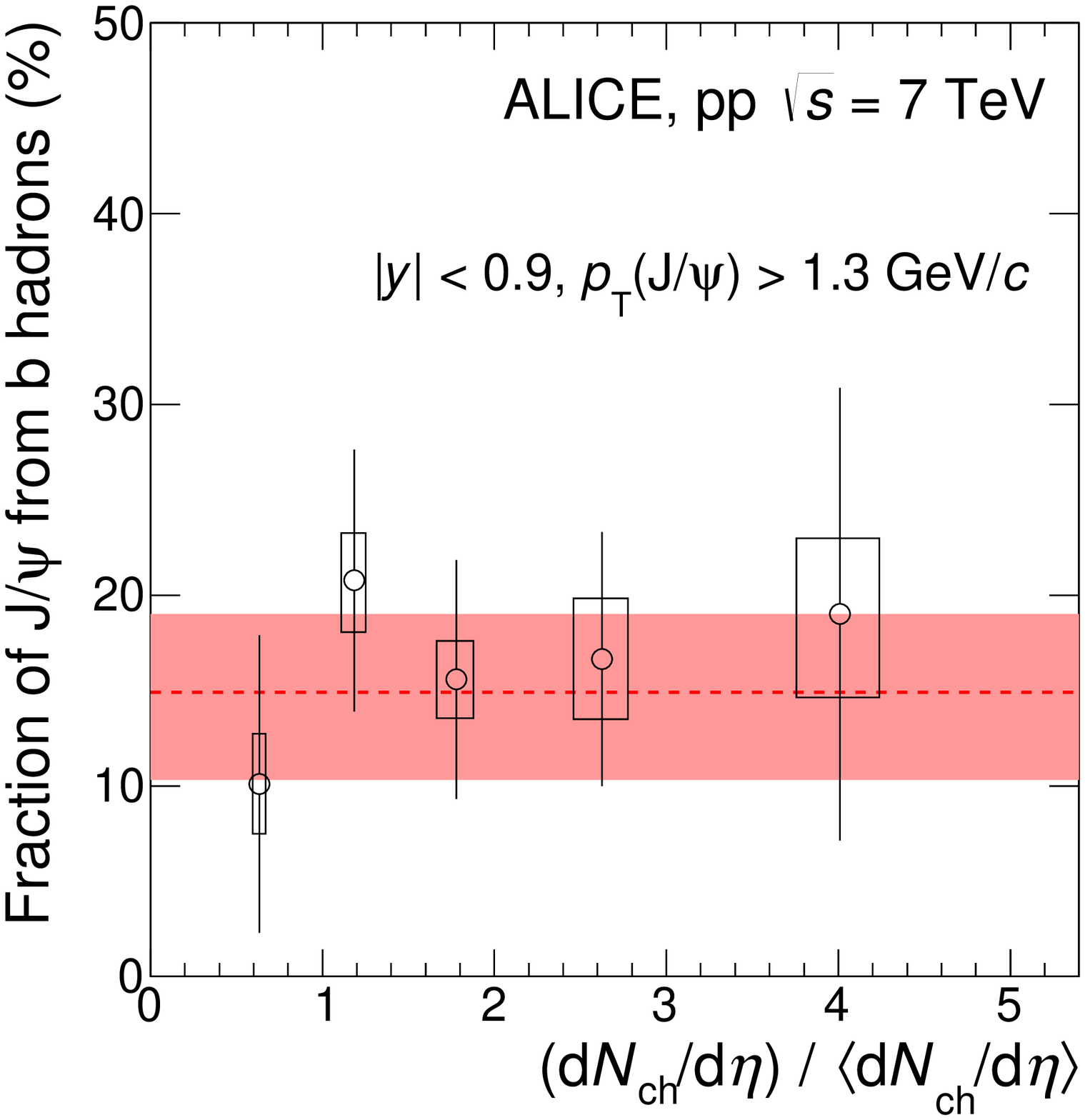}
\caption{Non-prompt $\Jpsi$ fraction as a function of the relative charged-particle multiplicity at central rapidity for $\pt >1.3~\gevc$. 
The vertical bars represent the statistical uncertainties, while the empty boxes stand for the systematic uncertainties. The width and the height of these empty boxes indicate the measurement uncertainty on the horizontal and vertical axis respectively. 
The dashed line shows the value of $\fB$ measured in the same $\pt$ range and integrated over multiplicity~\cite{Abelev:2012gx}. The shaded area represents the statistical and systematic uncertainties on the multiplicity-integrated result added in quadrature.
\label{fig:nonPromptJpsi}
}
\end{center}
\end{figure}

The results for the fraction of non-prompt $\Jpsi$ for both $\pt >0$ and $\pt >1.3~\gevc$, the relative yields of prompt and non-prompt $\Jpsi$ in each multiplicity bin for $\pt >0$ are summarized in Tables~\ref{tab:fbtableNoSigmaInel} and~\ref{tab:fbtable} and shown in Fig.~\ref{fig:nonPromptJpsi}.

\section{Comparison of charm and beauty production}
\label{sec:results}

\begin{figure}[!htbp]
\begin{center}
\subfigure[D meson with $2<\pt<4~\gev/c$ and inclusive $\Jpsi$]{
	\label{fig:DYieldCorrVsInclusiveJpsi}
	\includegraphics[width=0.475\columnwidth]{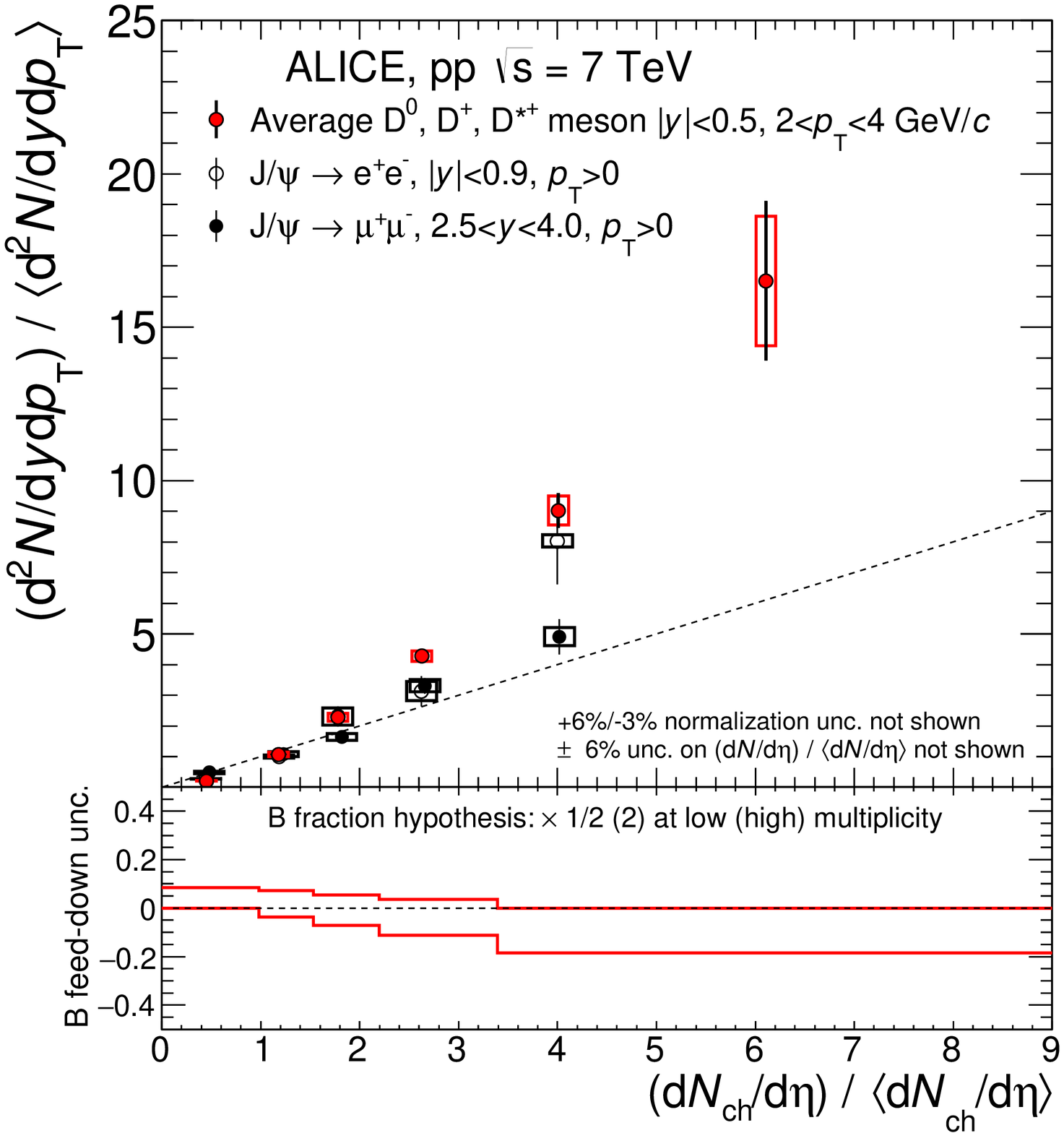}
	} 
\subfigure[D meson with $2<\pt<4~\gev/c$ and prompt $\Jpsi$]{
	\label{fig:DYieldCorrVsPromptJpsi}
	\includegraphics[width=0.475\columnwidth]{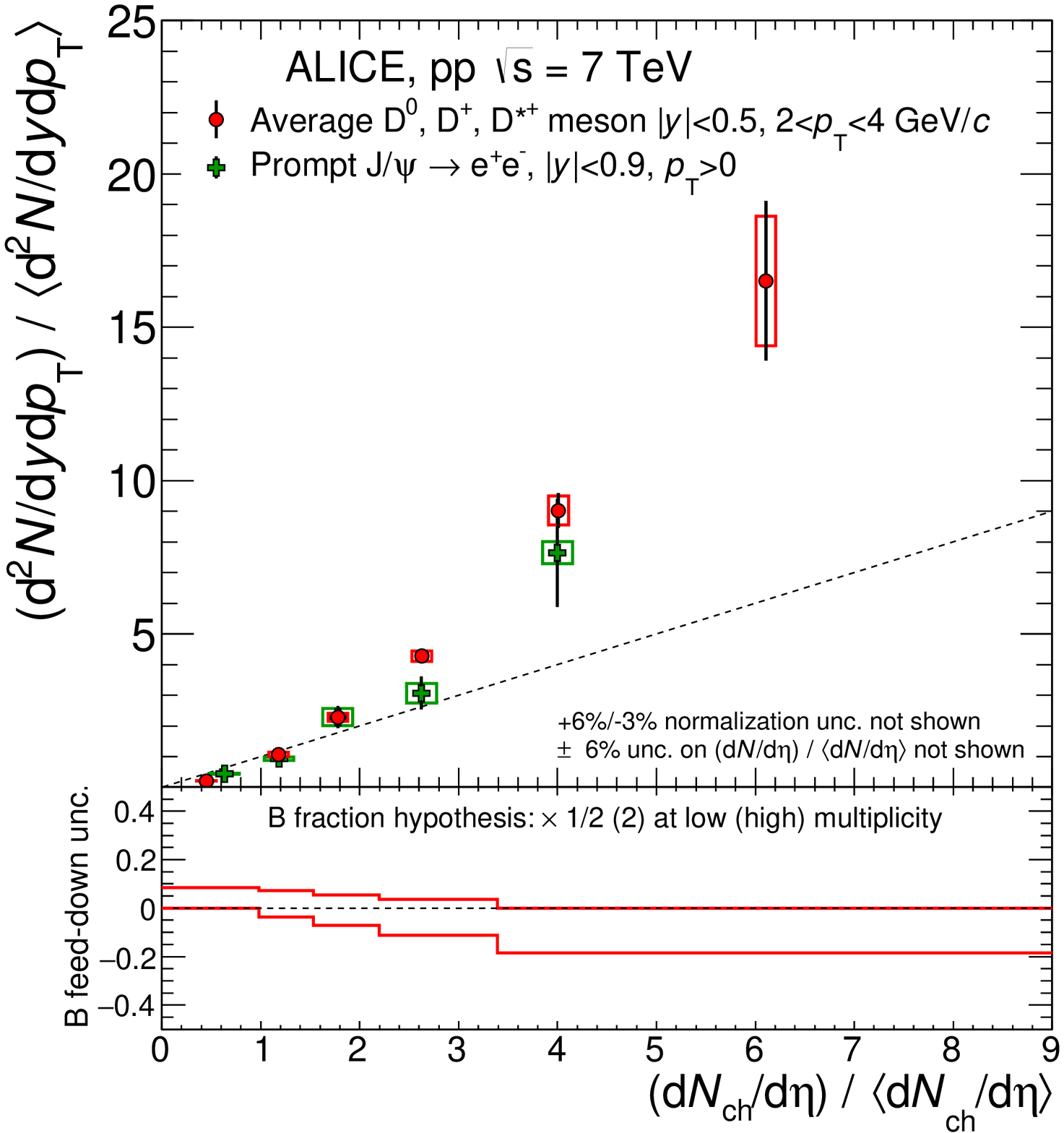}
	}
\subfigure[D meson with $2<\pt<4~\gev/c$ and non-prompt $\Jpsi$]{
	\label{fig:DYieldCorrVsNonPromptJpsi}
	\includegraphics[width=0.475\columnwidth]{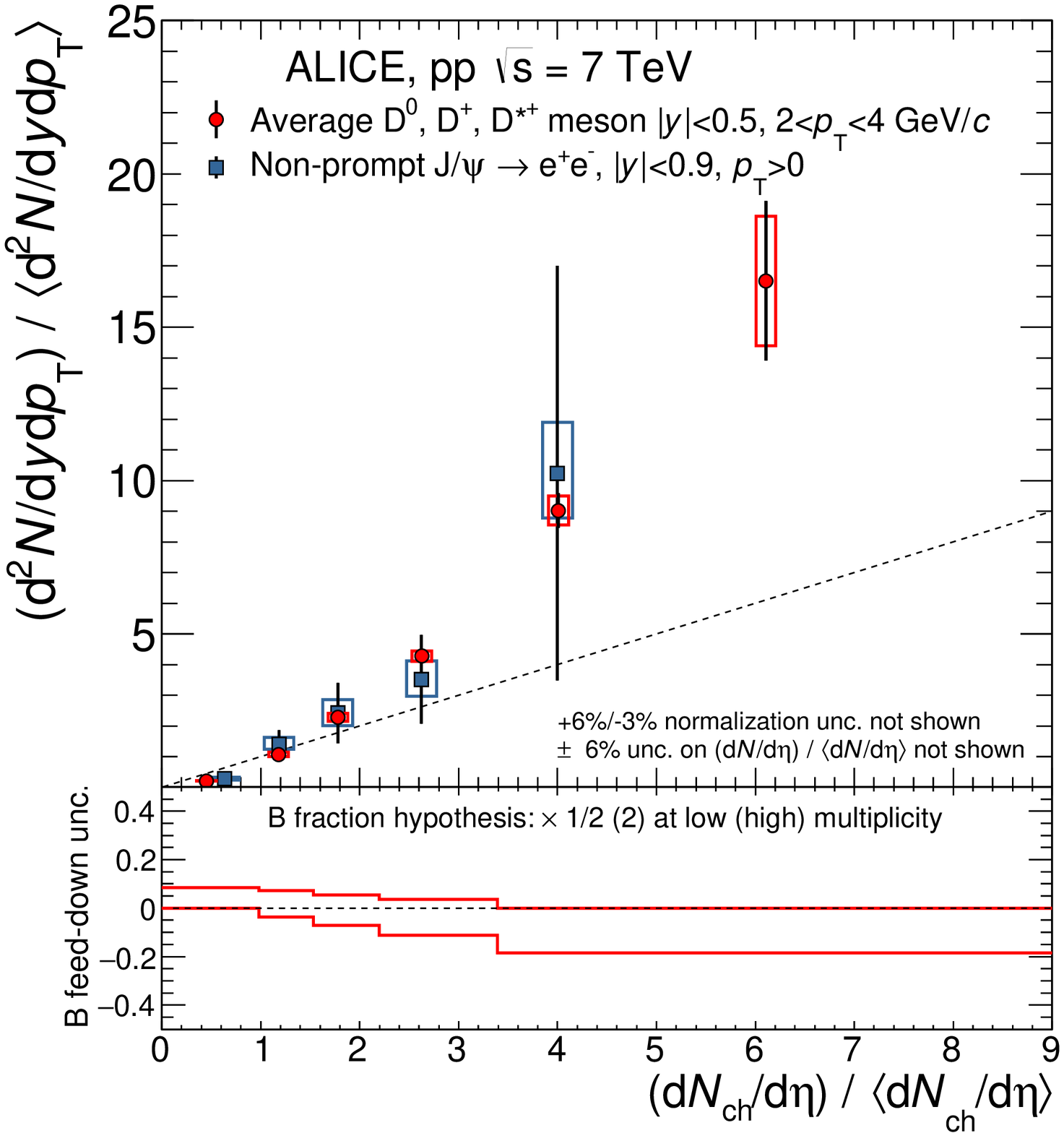}
	}
\caption{Average D meson and $\Jpsi$~relative yields as a function of the relative charged-particle multiplicity at central rapidity. D-meson yields are shown for $2<\pt<4~\gev/c$, while $\Jpsi$ yields are for $\pt>0$.
	(a) Inclusive $\Jpsi$ results for $|y|<0.9$ are represented by empty black circles~\cite{Abelev:2012rz}, inclusive $\Jpsi$ results for $2.5<y<4.0$ by black filled symbols~\cite{Abelev:2012rz}, and prompt D mesons by red filled circles.
	(b) Prompt $\Jpsi$ results for $|y|<0.9$ are represented by green filled crosses, and prompt D mesons by red filled circles.
	(c) Non-prompt $\Jpsi$ results for $|y|<0.9$ are represented by blue filled squares, and prompt D mesons by red filled circles.
	The relative yields are presented on the top panels with their statistical (vertical bars) and systematic (boxes) uncertainties except the uncertainty on the feed-down fraction for D mesons, which is drawn separately on the bottom panels. The points are located on the x-axis at the average value of $(\dNdEta) \big/ \langle \dNdEta \rangle$. 
The diagonal (dashed) line is drawn to guide the eye.
\label{fig:DYieldCorrVsJpsi}
}
\end{center}
\end{figure}

Figure~\ref{fig:DYieldCorrVsInclusiveJpsi} presents prompt D meson and inclusive $\Jpsi$ results to compare open and hidden charm production. The average prompt D-meson results are shown in the $2<\pt<4~\gev/c$ interval with the $\pt$-integrated inclusive $\Jpsi$ measurement\footnote{
After the inclusive $\Jpsi$ measurement was published in reference~\cite{Abelev:2012rz}, there was an improvement of the ALICE measurement of the inelastic cross section in \pp~collisions at $\sqrts=7$~TeV. 
The improved evaluation of the inelastic cross section does not rely on Monte Carlo, hence the systematic uncertainty is larger~\cite{Abelev:2012sea}. 
To allow a proper comparison with the results reported here, we updated the published inclusive $\Jpsi$ measurement by the corresponding change of the trigger efficiency for inelastic collisions 0.864/0.85. The normalisation uncertainties were also changed from $1.5\%$ to $^{+6}_{-3}\%$. 
} at central and forward-rapidity by the ALICE experiment~\cite{Abelev:2012rz}. 
The results for prompt $\Jpsi$ at central rapidity from this paper ($\pt>0$) and for prompt D mesons ($2<\pt<4~\gev/c$) are compared in Fig.~\ref{fig:DYieldCorrVsPromptJpsi}. 
A similar increase of the relative yield with the charged-particle multiplicity is observed for open and hidden charm production both at central and forward rapidities.

Figure~\ref{fig:DYieldCorrVsNonPromptJpsi} superimposes the open charm and beauty production measurements reported in this paper showing the average prompt D-meson results in the $2<\pt<4~\gev/c$ interval and the $\pt$-integrated non-prompt $\Jpsi$ measurement at central rapidity. 
The results are compatible within the measurement uncertainties. 

Open charm, open beauty and hidden charm hadron relative yields present a similar increase with charged-particle multiplicity. 
The comparison of open and hidden heavy flavour production suggests that this behaviour is most likely related to the \ccbar and \bbbar production processes, and is not significantly influenced by hadronisation. 
The enhancement of the heavy-flavour relative yields with the charged-particle multiplicity is qualitatively consistent with the calculations of the contribution from MPIs to particle production at LHC energies~\cite{Bartalini:2010su,Sjostrand:1987su,Porteboeuf:2010dw}. 
It could also be explained by the naive picture that processes with large momentum exchange might be associated to a larger amount of gluon-radiation at LHC energies, but no specific model implementation of this effect exists yet. 
The comparison of the results with model calculations is shown in the next section.

\section{Comparison to theoretical calculations}
\label{sec:MCcalculations}
Figures~\ref{fig:DYieldCorrAverage},~\ref{fig:nonPromptJpsi} and~\ref{fig:DYieldCorrVsJpsi} evidence a correlation between heavy-flavour and charged-particle multiplicities. Heavy-flavour production is dominated by hard processes, while charged-particle yields are associated to the soft momentum scale processes. 
It is then interesting to compare our results with calculations of event generators, designed to be as close as possible to real events in their description of the hard and soft components. Even though several event generators are available, few of them include heavy quarks in a consistent way. One of these is PYTHIA~\cite{Sjostrand:2006za,Sjostrand:2007gs}, which will be discussed in more detail
in Sec.~\ref{sec:pythia}. 
In Sec.~\ref{sec:theory} a comparison to PYTHIA~8~\cite{Sjostrand:2007gs}, to the EPOS~3~\cite{Drescher:2000ha,Werner:2013tya} event generator results and to a percolation model~\cite{Ferreiro:2012fb,Ferreiro:2015gea} calculation are presented.

\subsection{PYTHIA~8 simulations}
\label{sec:pythia}

PYTHIA~8~\cite{Sjostrand:2007gs} is the C++ successor of PYTHIA~6~\cite{Sjostrand:2006za}. 
One of the major improvements in PYTHIA~8  with respect to PYTHIA~6 concerns the treatment of the MPI scenario, where the c and b quarks can be involved in MPI $2 \to 2$ hard subprocesses. This model improvement is fundamental for an understanding of heavy-flavour production as a function of multiplicity, as MPI can contribute to the observed phenomena. Here  PYTHIA~8.157 simulations with the "SoftQCD" process selection including colour reconnection and diffractive processes~\footnote{
In this simulation, single- and double-diffractive processes contribute to about $20\%$ of the cross section. 
} are discussed, which will be referred to as PYTHIA~8.

Heavy-flavour production in PYTHIA~8 proceeds via four main mechanisms: 
{\it (i)} The {\it first (hardest) hard process}, where the initial c/b quarks originate from the first $2 \to 2$ hard process, mostly by gluon fusion ($gg \to {\rm c}\bar{\rm c}$) or involving a c/b sea-quark (e.g. ${\rm c}{\rm u} \to {\rm c}{\rm u}$). 
{\it (ii)} The subsequent hard processes in MPI, produced via the same mechanisms as the {\it first hard process} but in consecutive interactions, that we refer to as {\it hard process in MPI}.
Each produced gluon has a probability to split into a \ccbar or \bbbar pair contributing to heavy-flavour production. When the initial gluon originates from a hard process, either the first one or a subsequent process (in MPI), we refer to this process as {\it (iii)} {\it gluon splitting from hard process}. 
When the initial gluon originates from initial or final state radiation, we refer to this process as {\it (iv)} {\it ISR/FSR}. 

The contribution of the various production processes to the total D- and B-meson production in PYTHIA~8 for \pp~collisions at $\sqrts=7$~TeV is summarised in Table~\ref{tab:D_B_pythia8_origin}. 
In the following, D mesons refer to the average of $\Dzero$, $\Dplus$, and $\Dstar$, while B mesons represent the average of $\rm B^0$, $\rm B^+$, and $\rm B^{*+}$. 
Initial and final state radiation are the main contributors to open heavy-flavour production in PYTHIA~8, corresponding to $\sim62\%$ for D mesons and $\sim40\%$ for B mesons. 
MPI correspond to $\sim21\%$ ($\sim24\%$) of the D-meson (B-meson) production, while the first hard process is contributing $\sim11\%$ for D mesons and $\sim36\%$ for B mesons. 
It should be noted that in PYTHIA~8 the largest contribution to hard processes comes from c sea-quarks and not from gluon fusion~\cite{Gluck:1977zm}.
\begin{table}[!htb]
\centering
\begin{tabular}{ll|c c|cc}
\hline
\multicolumn{2}{c|}  {Origin of c and b quark content}            &   \multicolumn{2}{c|}  {D mesons}      & \multicolumn{2}{c}  {B mesons}       \\[1.01ex]
\hline
First hard process                                                                   &                                            &      11\%         &          &           36\% &    \\
                                                                                               &    gluon fusion                     &                      & 2\%     &                &     15\%           \\
                                                                                               & c/b sea                                   &                     & 9\%     &                 &     21\%             \\
Hard process in MPI                                                                &                                            &       21\%        &         &              24\%  & \\                              
\multicolumn{2}{l|}{Gluon splitting  from hard process}                                                      &  6\%            &           &       included in ISR/FSR   &      \\
ISR/FSR                                                                                  &                                            &   62\%            &         &     40\%          & \\
Remnant                                                                                  &                                            &      $<0.2\%$    &          &   $<0.4\%$        & \\             
\hline 
\end{tabular}
\caption{\label{tab:D_B_pythia8_origin}
Contribution of the different production processes to the total D- and B-meson production in PYTHIA~8.157~\cite{Sjostrand:2007gs} for \pp~collisions at $\sqrts=7$~TeV.
}
\end{table}

\begin{figure}[!htb]
\begin{center}
\includegraphics[width=0.75\columnwidth]{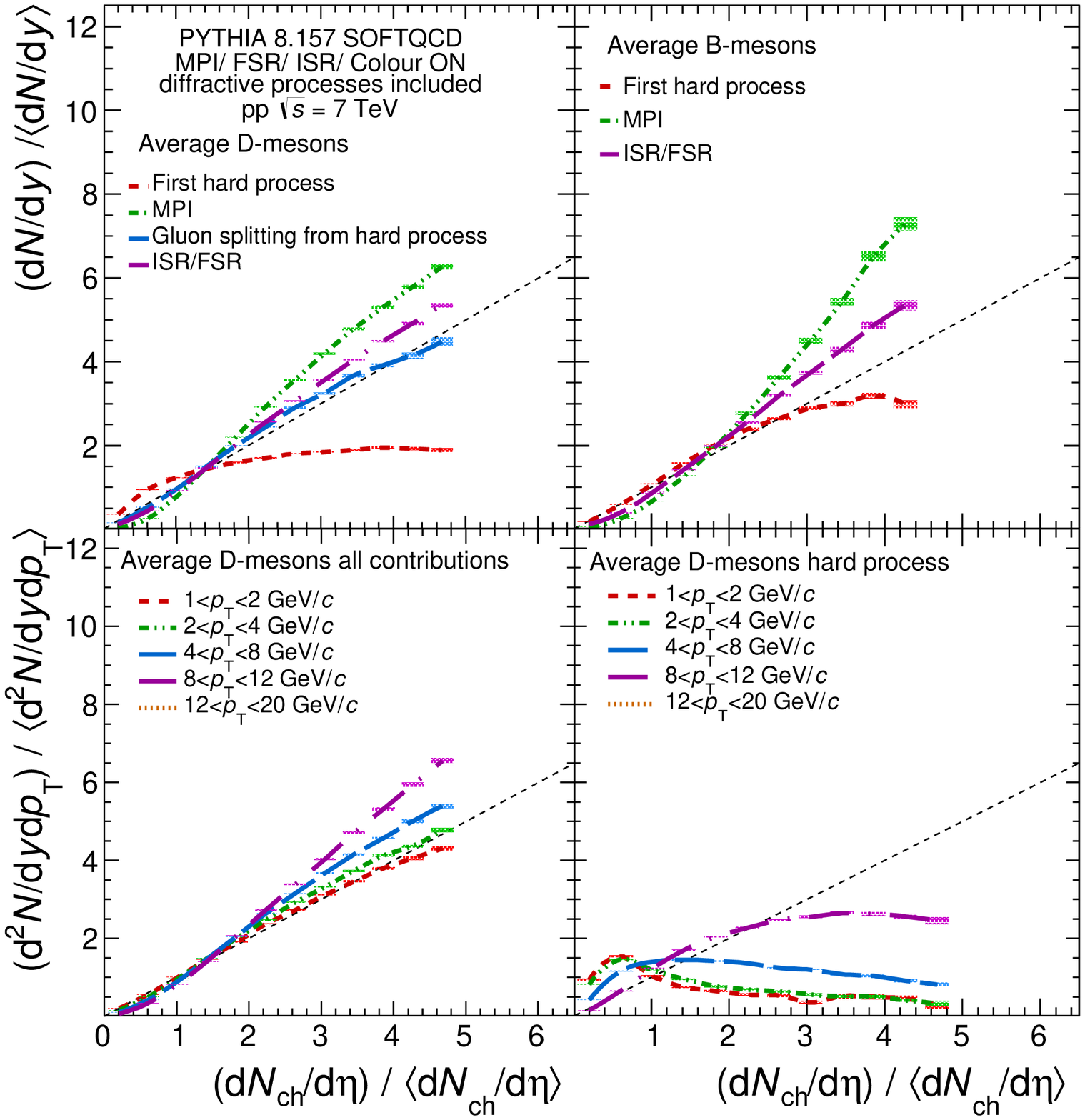}
\caption{
\label{fig:D_B_pythia8_origin}
D- and B-meson relative yield as a function of the relative charged-particle multiplicity at central rapidity calculated with the PYTHIA 8.157 event generator~\cite{Sjostrand:2007gs}. 
The different c and b quark production processes are separated on the top panels: first hard process, hard process in multiple interactions (MPI), gluon splitting from hard processes and initial/final state radiation (ISR/FSR). The bottom panels present the multiplicity dependence in several $\pt$ intervals for prompt D-meson production, on the left for all contributions and on the right for first hard process only.
The coloured lines represent the calculation distributions, whereas the shaded bands represent their statistical uncertainties at given values of $(\dNdEta) \big/ \langle \dNdEta \rangle$. 
The diagonal (dashed) line is drawn to guide the eye.
}
\end{center}
\end{figure}
Figure~\ref{fig:D_B_pythia8_origin} (top panels) shows the D and B-meson production as a function of the relative charged-particle multiplicity calculated with PYTHIA~8. The distributions for the main production processes are shown independently. The top-left panel presents results for D mesons, revealing an increasing trend of the relative yields as a function of the relative charged-particle multiplicity for MPI, the gluon splitting from hard processes, and the ISR/FSR contributions.
This is consistent with the fact that in PYTHIA~8 MPI and ISR/FSR contribute both to the total multiplicity and to heavy-flavour production. 
The first hard process contribution instead shows a weaker dependence on 
the multiplicity: a slight increase is observed at low multiplicities 
($\dNdEta \big/ \langle \dNdEta \rangle < 1$) followed by a saturation.
The picture for B mesons, on the top-right panel, presents similar features as that of D mesons.
The trend for the first hard process contribution shows an increase at low multiplicities and then saturates. The relative charged-particle multiplicity at which the plateau sets in is higher for B than for D mesons. The other contributions to particle production increase faster with multiplicity for B than for D mesons. These differences can be understood as being due to the larger B-meson mass, allowing a larger event activity in MPI and ISR/FSR processes.

Figure~\ref{fig:D_B_pythia8_origin} (bottom panels) presents the D-meson relative yields as a function of the relative charged-particle multiplicity in PYTHIA~8 for five $\pt$ intervals. The bottom-left panel shows the trend for the sum of all contributions, where an overall linear behaviour is observed, the slope of which increases with $\pt$.
The bottom-right panel shows the $\pt$ evolution for the first hard processes only. The relative D-meson yield decreases with multiplicity at low $\pt$ ($1<\pt<2~\gevc$), while at high $\pt$ ($12<\pt<20~\gevc$) it exhibits a linear increase. 
This feature is caused in PYTHIA~8 by the fact that MPI are ordered by their hardness, i.e. the $\pt$ of the first hard scattering is an upper limit for the subsequent hard scatterings and the related ISR/FSR. 
Thus, charm and beauty production at low $\pt$ is associated mostly with 
low multiplicity events, whereas heavy-flavour hadron production in high $\pt$ 
intervals is associated to higher multiplicity events.
For completeness, the contribution of MPI to the total charged-particle multiplicity was studied. Only events with a small number of MPI contribute to the low multiplicity intervals, while high multiplicity events are dominated by a large number of MPI, e.g. events with about five times the average multiplicity can have more than 16 parton--parton interactions in the event.

In the following, the multiplicity dependence for D and B-meson production including all contributions in a given $\pt$ interval, as shown in Fig.~\ref{fig:D_B_pythia8_origin} (bottom-left panel) for D-mesons, is compared to the measurements.

\subsection{Comparison of data with models}
\label{sec:theory}

\begin{figure}[!htb]
\begin{center}
\includegraphics[width=0.75\columnwidth]{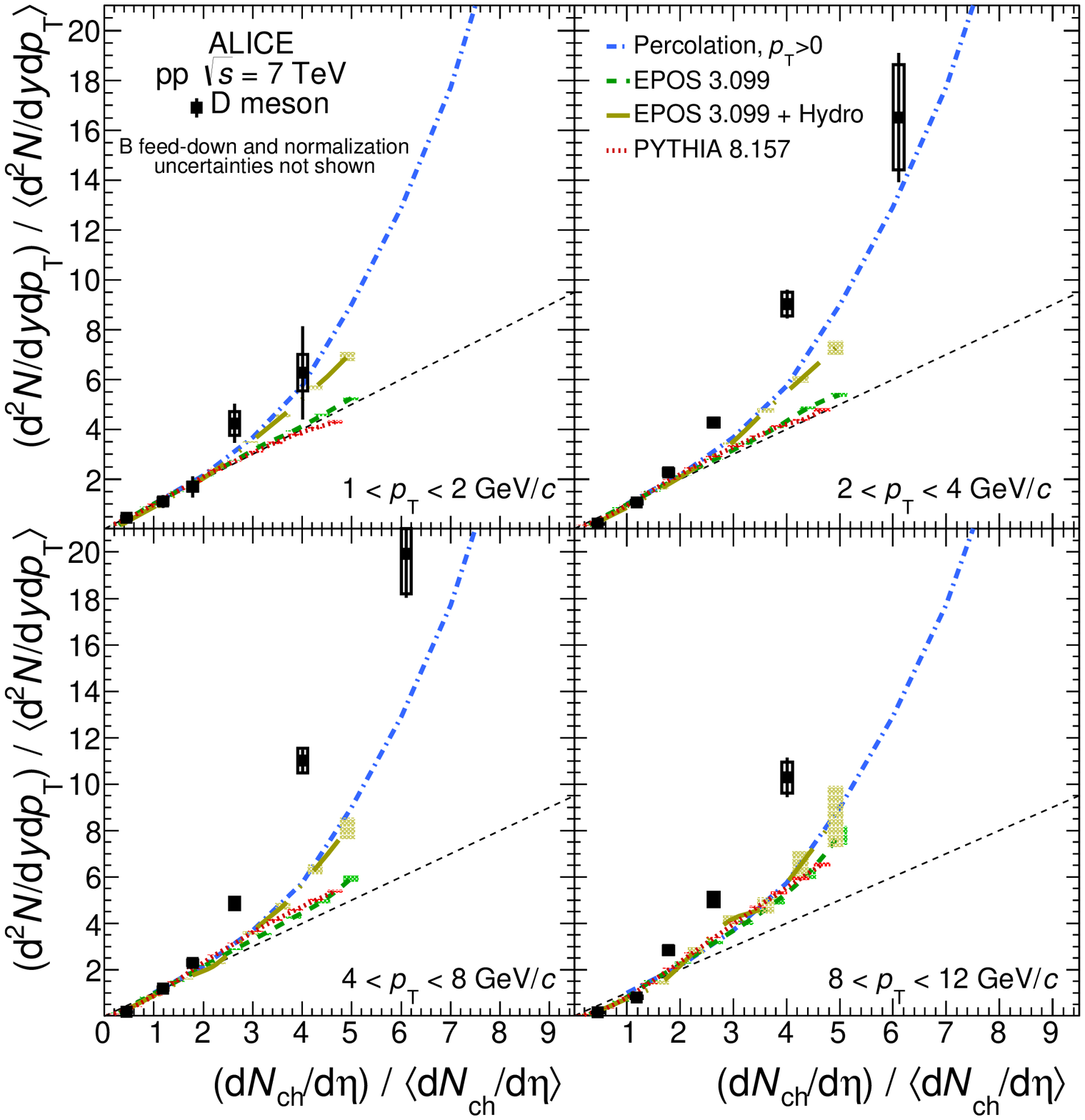}
\caption{
\label{fig:D_theory_summary}
Average D-meson relative yield as a function of the relative charged-particle multiplicity at central rapidity in different $\pt$ intervals. 
The systematic uncertainties on the data normalisation ($+6\%/-3\%$), on the $(\dNdEta) \big/ \langle \dNdEta \rangle$ values ($\pm6\%$), and on the feed down contribution are not shown in this figure.
Different calculations are presented: PYTHIA~8.157~\cite{Sjostrand:2006za,Sjostrand:2007gs}, EPOS~3 with and without hydro~\cite{Drescher:2000ha,Werner:2013tya} and a $\pt$-integrated percolation model~\cite{Ferreiro:2012fb,Ferreiro:2015gea}. 
The coloured lines represent the calculation curves, whereas the shaded bands represent their statistical uncertainties at given values of $(\dNdEta) \big/ \langle \dNdEta \rangle$. 
The diagonal (dashed) line is shown to guide the eye.
}
\end{center}
\end{figure}

\begin{figure}[!htb]
\begin{center}
\includegraphics[width=0.45\columnwidth]{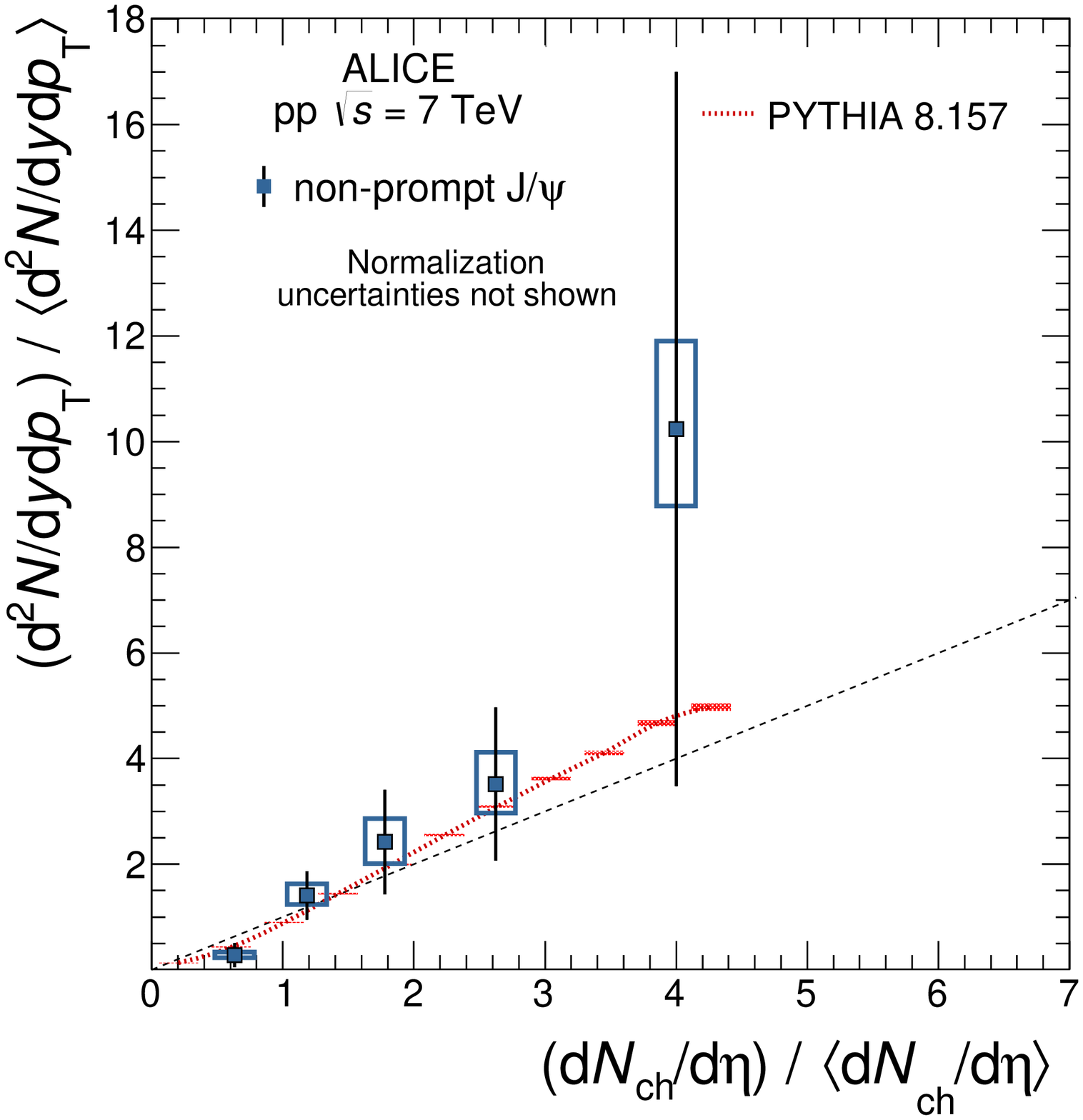}
\caption{
\label{fig:B_theory_summary}
Non-prompt $\Jpsi$ relative yield as a function of the relative charged-particle multiplicity at central rapidity for $\pt>0$. 
The systematic uncertainties on the data normalisation ($+6\%/-3\%$) and on the $(\dNdEta) \big/ \langle \dNdEta \rangle$ values ($\pm6\%$) are not shown in this figure. 
PYTHIA~8.157~\cite{Sjostrand:2007gs} calculation for B mesons is also presented. 
The coloured line represents the calculation curve, whereas the shaded band represents its uncertainty at given values of $(\dNdEta) \big/ \langle \dNdEta \rangle$. 
The diagonal (dashed) line is shown to guide the eye.
}
\end{center}
\end{figure}

Figure~\ref{fig:D_theory_summary} shows the comparison between D-meson (average of $\Dzero$, $\Dplus$ and $\Dstar$) production and theoretical calculations in four $\pt$ intervals. The results of the PYTHIA 8~\cite{Sjostrand:2006za,Sjostrand:2007gs} and the EPOS~3~\cite{Drescher:2000ha,Werner:2013tya} event generators, and of percolation calculations~\cite{Ferreiro:2012fb,Ferreiro:2015gea} are represented by the red dotted line, green dashed or long-dashed and dotted line, and the blue dot-dashed line, respectively. 
The description of the PYTHIA 8 setup was discussed in Sec.~\ref{sec:pythia}.
Figure~\ref{fig:B_theory_summary} presents $\pt$-integrated non-prompt $\Jpsi$ results together with PYTHIA 8~\cite{Sjostrand:2006za,Sjostrand:2007gs} calculations.
The percolation model assumes that high-energy hadronic collisions are driven by the exchange of colour sources between the projectile and target in the collision~\cite{Ferreiro:2012fb,Ferreiro:2015gea}. These colour sources have a finite spatial extension and can interact. In a high-density environment, the coherence among the sources leads to a reduction of their effective number. The source transverse mass determines its transverse size ($\propto 1/m_{\rm T}$), and allows to distinguish between soft (light) and hard (heavy) sources. 
As a consequence, at high densities the total charged-particle multiplicity, which originates from soft sources, is reduced. In contrast, hard particle production is less affected due to the smaller transverse size of hard sources. 
The percolation model predicts a faster-than-linear increase of heavy flavour relative production with the relative charged-particle multiplicity. 
The D-meson $\pt$-integrated percolation calculation is represented in all panels of Fig.~\ref{fig:D_theory_summary}, even though in this scenario a $\pt$ dependence of the results is expected, such that the higher the $\pt$ of the particle the stronger the deviation from the linear expectation.

EPOS~3~\cite{Drescher:2000ha,Werner:2013tya} is an event generator for various colliding systems: \pp, p--A and A--A. This event generator imposes the same theoretical scheme in all the systems, i.e. it assumes initial conditions followed by a hydrodynamical evolution. 
Initial conditions are generated in the Gribov-Regge multiple scattering framework, using the "Parton based Gribov-Regge" formalism~\cite{Drescher:2000ha}. Individual scatterings are referred to as Pomerons, and are identified with parton ladders. Each parton ladder is composed of a pQCD hard process with initial and final state radiation. Non-linear effects are considered by means of a saturation scale. The hadronisation is performed with a string fragmentation procedure. 
Based on these initial conditions, a hydrodynamical evolution can be applied on the dense core of the collision (3+1D viscous hydrodynamics)~\cite{Werner:2013tya}. 
An evaluation within the EPOS~3 model shows that the energy density reached in \pp~collisions at $\sqrts=7$ TeV is high enough to apply such hydrodynamic evolution~\cite{Werner:2013tya}. 
Here we discuss the results of an EPOS~3.099 calculation without jet-bulk interaction, which is a process that produces hadrons from hard partons and quarks from the fluid. 
EPOS~3 without the hydro component (green dashed line in Fig.~\ref{fig:D_theory_summary}) predicts an approximately linear increase of D-meson production as a function of the charged-particle multiplicity. This linear scaling shows a $\pt$ dependence, as observed in PYTHIA 8 with the colour reconnection scenario (red dotted line in Fig.~\ref{fig:D_theory_summary}), although the results differ in magnitude. 
In EPOS~3, a consequence of the Parton based Gribov-Regge approach is that the number of MPIs is directly related to the multiplicity, i.e. $N_{\rm hard \ process} \propto N_{\rm ch} \propto N_{\rm MPI}$.
When the hydrodynamic evolution is considered (green long-dashed and dotted line in Fig.~\ref{fig:D_theory_summary}), one observes a departure from a linear multiplicity dependence which is qualitatively comparable to that of the $\pt$-integrated percolation calculation.

The measurements, see Fig.~\ref{fig:D_theory_summary} and Fig.~\ref{fig:B_theory_summary}, provide evidence for an increase of the relative heavy-flavour yields with the relative charged-particle multiplicity which proceeds faster than linearly for high multiplicities. 
This result tends to favour calculations with a substantial deviation from linearity at high multiplicities such as EPOS~3.099 with hydrodynamics or the percolation model.

%
\section{Summary}
\label{sec:conclusions}
Charm and beauty hadron production as a function of the charged-particle multiplicity was studied in $\pp$~collisions at $\sqrts=7$~TeV. 
Charged-particle multiplicity at central rapidity was evaluated for events with at least a charged particle in the interval $|\eta|<1.0$. 
Prompt $\Dzero$, $\Dplus$ and $\Dstar$ meson yields were measured at central rapidity ($|y|<0.5$) in their hadronic decay channels in five $\pt$ intervals, from $1~\gevc$ to $20~\gevc$. 
The increase of the relative yield with increasing charged-particle multiplicity was found to be similar for D-meson species in all investigated $\pt$ intervals.
The average of the $\Dzero$, $\Dplus$ and $\Dstar$ relative yields increase with the relative charged-particle multiplicity faster than linearly at high multiplicities. 
No $\pt$ dependence is observed within the current statistical and systematic uncertainties. 
The lack of quantitative model estimates of the $\pt$ dependence together with the measurement uncertainties prevent to conclude on a possible $\pt$ dependence. 
A relative yield enhancement of about a factor of 15 with respect to the multiplicity integrated value is observed for events with six times the average charged-particle multiplicity. 
Prompt $\Dzero$ relative yields were also measured as a function of the relative charged-particle multiplicity determined in the pseudo-rapidity intervals $-3.7 < \eta < -1.7$ and $2.8 < \eta < 5.1$. 
The results were found to be consistent with those obtained using the charged-particle multiplicity measured at central rapidity.
$\Jpsi$ inclusive yields were measured earlier at central rapidity ($|y|<0.9$) in their di-electron decay channel~\cite{Abelev:2012rz}. The non-prompt $\Jpsi$ contribution was evaluated for $\pt>1.3~\gevc$, and extrapolated to $\pt>0$. 
The non-prompt $\Jpsi$ fraction does not show a dependence on the charged-particle multiplicity at central rapidity. 

Open charm, open beauty, and hidden charm hadron yields exhibit a similar increase with the charged-particle multiplicity at central rapidity. This suggests that heavy-flavour relative yields enhancement is not significantly influenced by hadronisation, but more likely directly related to the \ccbar and \bbbar production processes.
The heavy-flavour relative yield enhancement as a function of the charged-particle multiplicity is qualitatively described by: 
{\it (a)} PYTHIA~8.157 calculations including the MPI contributions to particle production~\cite{Sjostrand:2007gs},
{\it (b)} percolation model estimates of the influence of colour charge exchanges during the interaction~\cite{Ferreiro:2012fb,Ferreiro:2015gea}, 
{\it (c)} predictions by the EPOS~3 event generator which provides a description of the initial conditions followed by a hydrodynamical evolution~\cite{Drescher:2000ha,Werner:2013tya}. 
However, the PYTHIA~8.157~\cite{Sjostrand:2007gs} event generator seems to under-estimate the increase of heavy flavour yields with the charged-particle multiplicity at high multiplicities. 

%

\newenvironment{acknowledgement}{\relax}{\relax}
\begin{acknowledgement}
\section*{Acknowledgements}
We would like to thank P.~Skands, co-author of PYTHIA 8,  
and K.~Werner and B.~Guiot, co-authors of EPOS~3, for fruitful discussions and for providing their theoretical calculations.

The ALICE Collaboration would like to thank all its engineers and technicians for their invaluable contributions to the construction of the experiment and the CERN accelerator teams for the outstanding performance of the LHC complex.
The ALICE Collaboration gratefully acknowledges the resources and support provided by all Grid centres and the Worldwide LHC Computing Grid (WLCG) collaboration.
The ALICE Collaboration acknowledges the following funding agencies for their support in building and
running the ALICE detector:
State Committee of Science,  World Federation of Scientists (WFS)
and Swiss Fonds Kidagan, Armenia,
Conselho Nacional de Desenvolvimento Cient\'{\i}fico e Tecnol\'{o}gico (CNPq), Financiadora de Estudos e Projetos (FINEP),
Funda\c{c}\~{a}o de Amparo \`{a} Pesquisa do Estado de S\~{a}o Paulo (FAPESP);
National Natural Science Foundation of China (NSFC), the Chinese Ministry of Education (CMOE)
and the Ministry of Science and Technology of China (MSTC);
Ministry of Education and Youth of the Czech Republic;
Danish Natural Science Research Council, the Carlsberg Foundation and the Danish National Research Foundation;
The European Research Council under the European Community's Seventh Framework Programme;
Helsinki Institute of Physics and the Academy of Finland;
French CNRS-IN2P3, the `Region Pays de Loire', `Region Alsace', `Region Auvergne' and CEA, France;
German Bundesministerium fur Bildung, Wissenschaft, Forschung und Technologie (BMBF) and the Helmholtz Association;
General Secretariat for Research and Technology, Ministry of
Development, Greece;
Hungarian Orszagos Tudomanyos Kutatasi Alappgrammok (OTKA) and National Office for Research and Technology (NKTH);
Department of Atomic Energy and Department of Science and Technology of the Government of India;
Istituto Nazionale di Fisica Nucleare (INFN) and Centro Fermi -
Museo Storico della Fisica e Centro Studi e Ricerche "Enrico
Fermi", Italy;
MEXT Grant-in-Aid for Specially Promoted Research, Ja\-pan;
Joint Institute for Nuclear Research, Dubna;
National Research Foundation of Korea (NRF);
Consejo Nacional de Cienca y Tecnologia (CONACYT), Direccion General de Asuntos del Personal Academico(DGAPA), M\'{e}xico, Amerique Latine Formation academique - European Commission~(ALFA-EC) and the EPLANET Program~(European Particle Physics Latin American Network);
Stichting voor Fundamenteel Onderzoek der Materie (FOM) and the Nederlandse Organisatie voor Wetenschappelijk Onderzoek (NWO), Netherlands;
Research Council of Norway (NFR);
National Science Centre, Poland;
Ministry of National Education/Institute for Atomic Physics and National Council of Scientific Research in Higher Education~(CNCSI-UEFISCDI), Romania;
Ministry of Education and Science of Russian Federation, Russian
Academy of Sciences, Russian Federal Agency of Atomic Energy,
Russian Federal Agency for Science and Innovations and The Russian
Foundation for Basic Research;
Ministry of Education of Slovakia;
Department of Science and Technology, South Africa;
Centro de Investigaciones Energeticas, Medioambientales y Tecnologicas (CIEMAT), E-Infrastructure shared between Europe and Latin America (EELA), Ministerio de Econom\'{i}a y Competitividad (MINECO) of Spain, Xunta de Galicia (Conseller\'{\i}a de Educaci\'{o}n),
Centro de Aplicaciones Tecnológicas y Desarrollo Nuclear (CEA\-DEN), Cubaenerg\'{\i}a, Cuba, and IAEA (International Atomic Energy Agency);
Swedish Research Council (VR) and Knut $\&$ Alice Wallenberg
Foundation (KAW);
Ukraine Ministry of Education and Science;
United Kingdom Science and Technology Facilities Council (STFC);
The United States Department of Energy, the United States National
Science Foundation, the State of Texas, and the State of Ohio;
Ministry of Science, Education and Sports of Croatia and  Unity through Knowledge Fund, Croatia.
Council of Scientific and Industrial Research (CSIR), New Delhi, India
\end{acknowledgement}


\bibliographystyle{utphys}
\bibliography{DJpsivsMultpp.bib}

\appendix

\section{Tables of the results}

Table~\ref{tab:DAverage} reports the results of the relative average D-meson yields per inelastic collision as a function of the relative charged-particle multiplicity at mid-rapidity in several D-meson transverse momentum intervals, see Fig.~\ref{fig:DYieldCorrAverage}. The corresponding relative average D-meson yields normalised to the visible cross section instead of the inelastic one are presented in Table~\ref{tab:DAverageNoSigmaInel}. 

Table~\ref{tab:DvsNvzero} summarises the relative $\Dzero$ yields per inelastic collision as a function of the relative raw multiplicity measured with the V0 detector at forward rapidity, see Fig.~\ref{fig:DzeroYieldCorrVzero}. These relative $\Dzero$ yields are also presented in Table~\ref{tab:DvsNvzeroNoSigmaInel} normalised to the visible cross section. 

Table~\ref{tab:fbtableNoSigmaInel} reports the fraction of non-prompt $\Jpsi$ to the inclusive $\Jpsi$ yields as a function of the relative charged-particle multiplicity at mid-rapidity, see Fig~\ref{fig:nonPromptJpsi}. The relative prompt and non-prompt $\Jpsi$ yields per inelastic collision are reported in Table~\ref{tab:fbtable}, while Table~\ref{tab:fbtableNoSigmaInel} presents these yields normalised to the visible cross section.

\begin{landscape}

\begin{table}[!htdp]
{\footnotesize
\begin{center}\begin{tabular}{lcccccc}
\hline
 &  \multicolumn{6}{c}{$\Ntrk$ interval} \\
& [1,8] & [9,13] & [14,19] & [20,30] & [31,49] & [50,80] \\[1.01ex] \hline
$\pt$~($\gevc$)&  \multicolumn{6}{c}{$({\rm d}^2N/{\rm d}y{\rm d}p_{\rm T}) \big/ \langle {\rm d}^2N/{\rm d}y{\rm d}p_{\rm T} \rangle  \times \epsilon_{\rm trigger}$} \\[1.01ex]
1--2 &
	$0.39 \pm 0.09 \pm 0.05 _{-0}^{+0.07}$ &
	$0.94 \pm 0.21 \pm 0.11 _{-0.07}^{+0.14}$ &
	$1.45 \pm 0.36 \pm 0.15 _{-0.22}^{+0.16}$ &
	$3.60 \pm 0.66 \pm 0.40 _{-0.82}^{+0.27}$ &
	$5.33 \pm 1.59 \pm 0.62 _{-2.01}^{+0}$ &
	 --\\[1.01ex]%
2--4 & 
	$0.18 \pm 0.01 \pm 0.01 _{-0}^{+0.02}$ &
	$0.91 \pm 0.05 \pm 0.06 _{-0.03}^{+0.07}$ &
	$1.94 \pm 0.10 \pm 0.11 _{-0.14}^{+0.10}$ &
	$3.64 \pm 0.16 \pm 0.15 _{-0.40}^{+0.13}$ &
	$7.67 \pm 0.48 \pm 0.40 _{-1.42}^{+0}$ &
	$14.04 \pm 2.21 \pm 1.79 _{-3.28}^{+0}$ \\[1.01ex]
4--8 & %
	$0.15 \pm 0.01 \pm 0.01 _{-0}^{+0.01}$ &
	$1.01 \pm 0.04 \pm 0.06 _{-0.03}^{+0.06}$ &
	$1.95 \pm 0.08 \pm 0.10 _{-0.12}^{+0.09}$ &
	$4.13 \pm 0.12 \pm 0.22 _{-0.35}^{+0.12}$ &
	$9.36 \pm 0.39 \pm 0.45 _{-1.49}^{+0}$ &
	$16.94 \pm 1.60 \pm 1.45 _{-3.15}^{+0}$ \\[1.01ex]
8--12 & 
	$0.14 \pm 0.02 \pm 0.01 _{-0}^{+0.01}$ &
	$0.69 \pm 0.07 \pm 0.05 _{-0.02}^{+0.04}$ &
	$2.41 \pm 0.15 \pm 0.15 _{-0.15}^{+0.11}$ &
	$4.27 \pm 0.25 \pm 0.26 _{-0.42}^{+0.14}$ &
	$8.74 \pm 0.72 \pm 0.57 _{-1.48}^{+0}$ &
--\\[1.01ex]
12--20 & -- & --	&
	$2.14 \pm 0.30 \pm 0.26 _{-0.10}^{+0.07}$ &
	$4.15 \pm 0.51 \pm 0.43 _{-0.28}^{+0.09}$ &
	$11.60 \pm 1.59 \pm 1.18 _{-1.31}^{+0}$ &
	 --\\[1.01ex]
\hline
\end{tabular}
\caption{Average of $\Dzero$, $\Dplus$ and $\Dstar$ mesons relative yields for the sum of particle and antiparticle in several multiplicity and $\pt$ intervals for \pp~collisions at~$\sqrt{s}=7$~TeV as a function of the relative charged-particle multiplicity at central rapidity. The values are reported together with their uncertainties, which are quoted in the the order: statistical, systematic and feed-down contribution uncertainties. The yields reported here are not corrected by the trigger selection efficiency, they are normalised to the visible cross section.
\label{tab:DAverageNoSigmaInel}
}

\end{center}
}
\end{table}

\begin{table}[!htdp]
{\footnotesize
\begin{center}

\begin{tabular}{lcccccc}
\hline
 &  \multicolumn{6}{c}{$(\dNdEta) \big/ \langle \dNdEta \rangle$} \\[1.01ex]
&  $0.45^{+0.03}_{-0.03}$ & $1.18^{+0.07}_{-0.07}$ & $1.78^{+0.10}_{-0.11}$ & $2.63^{+0.15}_{-0.17}$ & $4.01^{+0.23}_{-0.25}$ & $6.11^{+0.35}_{-0.39}$ \\[1.01ex]
\hline
$\pt$~($\gevc$)&  \multicolumn{6}{c}{$({\rm d}^2N/{\rm d}y{\rm d}p_{\rm T}) \big/ \langle {\rm d}^2N/{\rm d}y{\rm d}p_{\rm T} \rangle $} \\[1.01ex]
1--2 & $0.45 \pm 0.11 \pm 0.05 _{-0}^{+0.09}$ &
	  $1.11 \pm 0.25 \pm 0.13 _{-0.08}^{+0.17}$ &
	  $1.70 \pm 0.43 \pm 0.18 _{-0.26}^{+0.19}$ &
	  $4.24 \pm 0.78 \pm 0.47 _{-0.96}^{+0.32}$ &
	  $6.27 \pm 1.87 \pm 0.73 _{-2.37}^{+0}$ &
	 --\\[1.01ex]
2--4 & $0.21 \pm 0.02 \pm 0.01 _{-0}^{+0.02}$ &
	$1.08 \pm 0.06 \pm 0.06 _{-0.04}^{+0.08}$ &
	$2.28 \pm 0.12 \pm 0.13 _{-0.16}^{+0.12}$ &
	$4.28 \pm 0.19 \pm 0.17 _{-0.48}^{+0.16}$ &
	$9.02 \pm 0.57 \pm 0.47 _{-1.67}^{+0}$ &
	$16.51 \pm 2.60 \pm 2.11 _{-3.86}^{+0}$ \\[1.01ex]
4--8 & $0.18 \pm 0.01 \pm 0.01 _{-0}^{+0.01}$ &
	$1.18 \pm 0.05 \pm 0.07 _{-0.04}^{+0.07}$ &
	$2.30 \pm 0.09 \pm 0.12 _{-0.14}^{+0.11}$ &
	$4.85 \pm 0.15 \pm 0.26 _{-0.42}^{+0.14}$ &
	$11.02 \pm 0.46 \pm 0.53 _{-1.75}^{+0}$ &
	$19.92 \pm 1.89 \pm 1.71 _{-3.71}^{+0}$ \\[1.01ex]
8--12 & $0.16 \pm 0.02 \pm 0.01 _{-0.00}^{+0.01}$ &
	$0.81 \pm 0.08 \pm 0.06 _{-0.03}^{+0.05}$ &
	$2.84 \pm 0.17 \pm 0.18 _{-0.17}^{+0.13}$ &
	$5.02 \pm 0.29 \pm 0.31 _{-0.50}^{+0.17}$ &
	$10.28 \pm 0.85 \pm 0.67 _{-1.74}^{+0}$&
	 --\\[1.01ex]
12--20 & -- & --	&
	$2.52 \pm 0.35 \pm 0.30 _{-0.11}^{+0.09}$ &
	$4.88 \pm 0.61 \pm 0.51 _{-0.33}^{+0.11}$ &
	$13.65 \pm 1.87 \pm 1.38 _{-1.54}^{+0}$ &
	 --\\[1.01ex]
\hline
\end{tabular}
\caption{Average of $\Dzero$, $\Dplus$ and $\Dstar$ mesons relative yields for the sum of particle and antiparticle in several multiplicity and $\pt$ intervals for \pp~collisions at~$\sqrt{s}=7$~TeV as a function of the relative charged-particle multiplicity at central rapidity. The values are reported together with their uncertainties, which are quoted in the the order: statistical, systematic and feed-down contribution uncertainties. The yields reported here are per inelastic event.
\label{tab:DAverage}
}

\end{center}
}
\end{table}

\end{landscape}

\begin{landscape}

\begin{table}[htdp]
\begin{center}
\begin{tabular}{lccccc}
\hline
&  \multicolumn{5}{c}{$\Nvzero \big/ \langle \Nvzero \rangle$} \\[1.01ex]
& 0.43 & 1.0 & 1.5 & 2.2 & 3.3 \\
\hline
$\pt$~($\gevc$)&  \multicolumn{5}{c}{$({\rm d}^2N/{\rm d}y{\rm d}p_{\rm T}) \big/ \langle {\rm d}^2N/{\rm d}y{\rm d}p_{\rm T} \rangle  \times \epsilon_{\rm trigger}$} \\[1.01ex]
2--4 & 
	$0.15 \pm 0.02 \pm 0.01 ^{+0.02}_{-0}$ &
	$0.65 \pm 0.07 \pm 0.03 ^{+0.06}_{-0.03}$ &
	$1.28 \pm 0.11 \pm 0.07 ^{+0.09}_{-0.12}$ &
	$2.81 \pm 0.15 \pm 0.11 ^{+0.13}_{-0.39}$ &
	$4.22 \pm 0.33 \pm 0.24 ^{+0}_{-0.99}$ \\[1.01ex]
4--8 & 
	$0.17 \pm 0.02 \pm 0.01 ^{+0.02}_{-0}$ &
	$0.80 \pm 0.07 \pm 0.04 ^{+0.07}_{-0.04}$ &
	$1.50 \pm 0.12 \pm 0.08 ^{+0.10}_{-0.13}$ &
	$2.57 \pm 0.15 \pm 0.14 ^{+0.11}_{-0.34}$ &
	$4.53 \pm 0.34 \pm 0.24 ^{+0}_{-1.00}$  \\[1.01ex]
\hline
\end{tabular}
\caption{$\Dzero$ meson relative yields for the sum of particle and antiparticle in several multiplicity and $\pt$ intervals for \pp~collisions at~$\sqrt{s}=7$~TeV as a function of the relative average multiplicity in the V0 detector, $\Nvzero \big/ \langle \Nvzero \rangle$.
 The yields reported here are not corrected by the trigger selection efficiency, they are normalised to the visible cross section.
\label{tab:DvsNvzeroNoSigmaInel}
}

\end{center}
\end{table}

\begin{table}[htdp]
\begin{center}
\begin{tabular}{lccccc}
\hline
&  \multicolumn{5}{c}{$\Nvzero \big/ \langle \Nvzero \rangle$} \\[1.01ex]
& 0.43 & 1.0 & 1.5 & 2.2 & 3.3 \\
\hline
$\pt$~($\gevc$)&  \multicolumn{5}{c}{$({\rm d}^2N/{\rm d}y{\rm d}p_{\rm T}) \big/ \langle {\rm d}^2N/{\rm d}y{\rm d}p_{\rm T} \rangle $} \\[1.01ex]
2--4 & $0.18 \pm 0.02 \pm 0.01 ^{+0.02}_{-0}$ & 
	$0.76 \pm 0.08 \pm 0.04 ^{+0.07}_{-0.04}$ & 
	$1.50 \pm 0.13 \pm 0.08 ^{+0.11}_{-0.14}$ & 
	$3.31 \pm 0.18 \pm 0.13 ^{+0.15}_{-0.46}$ & 
	$4.96 \pm 0.38 \pm 0.28 ^{+0}_{-1.16}$ \\[1.01ex]
4--8 & $ 0.20 \pm 0.02 \pm 0.01 ^{+0.02}_{-0}$ & 
	$0.94 \pm 0.09 \pm 0.05 ^{+0.08}_{-0.04}$ & 
	$1.76 \pm 0.14 \pm 0.09 ^{+0.12}_{-0.16}$ & 
	$3.02 \pm 0.18 \pm 0.16 ^{+0.13}_{-0.40}$ & 
	$5.33 \pm 0.40 \pm 0.29 ^{+0}_{-1.18}$  \\[1.01ex]
\hline
\end{tabular}

\end{center}
\caption{$\Dzero$ meson relative yields for the sum of particle and antiparticle in several multiplicity and $\pt$ intervals for \pp~collisions at~$\sqrt{s}=7$~TeV as a function of the relative average multiplicity in the V0 detector, $\Nvzero \big/ \langle \Nvzero \rangle$.
The yields reported here are normalised to the inelastic cross section.
\label{tab:DvsNvzero}
}
\end{table}

\end{landscape}

\begin{table}[!htbp]
\centering
\resizebox{\columnwidth}{!}{
\begin{tabular}{ccccc}
\hline
$\Ntrk$ & $\fB$(\%) & $\fB^{\rm extr}$(\%) & 
$({\rm d}N_{\Jpsi}^{\rm prompt}/{\rm d}y) \big/ \langle {\rm d}N_{\Jpsi}^{\rm prompt}/{\rm d}y \rangle\times \epsilon_{\rm trigger}$ &
$({\rm d}N_{\Jpsi}^{\rm non-prompt}/{\rm d}y) \big/ \langle {\rm d}N_{\Jpsi}^{\rm non-prompt}/{\rm d}y \rangle\times \epsilon_{\rm trigger}$ \\[1.01ex]
\hline
[4,8]  & $10.1 \pm 7.8 \pm 2.5$ & $10.2 \pm 7.9 \pm 2.5$ & $ 0.37 \pm 0.07 \pm 0.01$ & $ 0.24 \pm 0.20 ^{+0.05}_{-0.04}$ \\[1.01ex]
[9,13] &  $20.8 \pm 6.9 \pm 2.7$ & $20.9 \pm 6.9 \pm 2.7$ & $ 0.80 \pm  0.14 \pm 0.04 $ & $ 1.20 \pm 0.39 ^{+0.19}_{-0.14}$ \\[1.01ex]
[14,19] & $15.6 \pm 6.3 \pm 2.0$ &  $15.7 \pm 6.3 \pm 2.0$  & $ 1.95 \pm 0.31 \pm 0.24 $ & $ 2.06 \pm 0.84 ^{+0.37}_{-0.35}$ \\[1.01ex]
[20,30] &  $16.7 \pm 6.7 \pm 3.3$ & $16.8 \pm 6.7 \pm 3.3$ & $ 2.61 \pm 0.46 \pm 0.27 $ & $ 2.99 \pm  1.23 ^{+0.51}_{-0.47}$ \\[1.01ex]
[31,49] &  $19.0 \pm 11.9 \pm 4.2$ & $19.0 \pm 12.0 \pm 4.2$ & $ 6.50 \pm 1.50 \pm 0.31 $ & $ 8.70 \pm 5.75 ^{+1.41}_{-1.24}$ \\[1.01ex]
\hline
\end{tabular}
}
\caption{Fraction of non-prompt $\Jpsi$ measured for $\pt >1.3~\gevc$, $\fB$(\%), and extrapolated down to $\pt >0$,  $\fB^{\rm extr}$(\%) in the various $\Ntrk$ intervals. 
Prompt and non-prompt $\Jpsi$ relative yields for $\pt >0$ are also reported in the different multiplicity intervals.
The first and second uncertainties correspond to the statistical and systematic uncertainties, respectively.
 The yields reported here are not corrected by the trigger selection efficiency, they are normalised to the visible cross section.
\label{tab:fbtableNoSigmaInel}
}
\end{table}

\begin{table}[!htbp]
\centering
\resizebox{\columnwidth}{!}{
\begin{tabular}{ccc}
\hline
$(\dNdEta) \big/ \langle \dNdEta \rangle$ &  
$({\rm d}N_{\Jpsi}^{\rm prompt}/{\rm d}y) \big/ \langle {\rm d}N_{\Jpsi}^{\rm prompt}/{\rm d}y \rangle$ &
$({\rm d}N_{\Jpsi}^{\rm non-prompt}/{\rm d}y) \big/ \langle {\rm d}N_{\Jpsi}^{\rm non-prompt}/{\rm d}y \rangle$ \\[1.01ex]
\hline
$0.63^{+0.4}_{-0.4}$  & $0.44 \pm 0.08 \pm 0.01$ & $0.28 \pm 0.23 ^{+0.06}_{-0.05}$ \\[1.01ex]
$1.18^{+0.07}_{-0.07}$ &  $0.94 \pm 0.17 \pm 0.05$ & $1.41 \pm 0.46 ^{+0.22}_{-0.17}$ \\[1.01ex]
$1.78^{+0.10}_{-0.11}$ &  $2.29 \pm 0.36 \pm 0.28$ & $2.42 \pm 0.99 ^{+0.44}_{-0.41}$ \\[1.01ex]
$2.63^{+0.15}_{-0.17}$ &  $3.07 \pm 0.54 \pm 0.32$ & $3.52 \pm 1.45 ^{+0.60}_{-0.55}$ \\[1.01ex]
$4.01^{+0.23}_{-0.25}$ &  $7.65 \pm 1.76 \pm 0.36$ & $10.24 \pm 6.76 ^{+1.66}_{-1.46}$ \\[1.01ex]
\hline
\end{tabular}
}
\caption{
Prompt and non-prompt relative $\Jpsi$ relative yields for $\pt >0$ in different multiplicity bins. 
The first and second uncertainties correspond to the statistical and systematic uncertainties respectively. 
The yields reported here are normalised to the inelastic cross section.
\label{tab:fbtable}
}

\end{table}

\newpage

\section{The ALICE Collaboration}
\label{app:collab}



\begingroup
\small
\begin{flushleft}
J.~Adam\Irefn{org40}\And
D.~Adamov\'{a}\Irefn{org83}\And
M.M.~Aggarwal\Irefn{org87}\And
G.~Aglieri Rinella\Irefn{org36}\And
M.~Agnello\Irefn{org111}\And
N.~Agrawal\Irefn{org48}\And
Z.~Ahammed\Irefn{org131}\And
I.~Ahmed\Irefn{org16}\And
S.U.~Ahn\Irefn{org68}\And
I.~Aimo\Irefn{org94}\textsuperscript{,}\Irefn{org111}
% ...
\renewcommand\labelenumi{\textsuperscript{\theenumi}~}

\section*{Affiliation notes}
\renewcommand\theenumi{\roman{enumi}}
\begin{Authlist}
\item \Adef{0}Deceased
\item \Adef{idp3748880}{Also at: M.V. Lomonosov Moscow State University, D.V. Skobeltsyn Institute of Nuclear, Physics, Moscow, Russia}
\item \Adef{idp5866224}{Also at: University of Kansas, Lawrence, Kansas, United States}
\end{Authlist}

\section*{Collaboration Institutes}
\renewcommand\theenumi{\arabic{enumi}~}
\begin{Authlist}

\item \Idef{org1}A.I. Alikhanyan National Science Laboratory (Yerevan Physics Institute) Foundation, Yerevan, Armenia
\item \Idef{org2}Benem\'{e}rita Universidad Aut\'{o}noma de Puebla, Puebla, Mexico
\item \Idef{org3}Bogolyubov Institute for Theoretical Physics, Kiev, Ukraine
\item \Idef{org4}Bose Institute, Department of Physics and Centre for Astroparticle Physics and Space Science (CAPSS), Kolkata, India
\item \Idef{org5}Budker Institute for Nuclear Physics, Novosibirsk, Russia
\item \Idef{org6}California Polytechnic State University, San Luis Obispo, California, United States
\item \Idef{org7}Central China Normal University, Wuhan, China
\item \Idef{org8}Centre de Calcul de l'IN2P3, Villeurbanne, France
\item \Idef{org9}Centro de Aplicaciones Tecnol\'{o}gicas y Desarrollo Nuclear (CEADEN), Havana, Cuba
\item \Idef{org10}Centro de Investigaciones Energ\'{e}ticas Medioambientales y Tecnol\'{o}gicas (CIEMAT), Madrid, Spain
\item \Idef{org11}Centro de Investigaci\'{o}n y de Estudios Avanzados (CINVESTAV), Mexico City and M\'{e}rida, Mexico
\item \Idef{org12}Centro Fermi - Museo Storico della Fisica e Centro Studi e Ricerche ``Enrico Fermi'', Rome, Italy
\item \Idef{org13}Chicago State University, Chicago, Illinois, USA
\item \Idef{org14}China Institute of Atomic Energy, Beijing, China
\item \Idef{org15}Commissariat \`{a} l'Energie Atomique, IRFU, Saclay, France
\item \Idef{org16}COMSATS Institute of Information Technology (CIIT), Islamabad, Pakistan
\item \Idef{org17}Departamento de F\'{\i}sica de Part\'{\i}culas and IGFAE, Universidad de Santiago de Compostela, Santiago de Compostela, Spain
\item \Idef{org18}Department of Physics and Technology, University of Bergen, Bergen, Norway
\item \Idef{org19}Department of Physics, Aligarh Muslim University, Aligarh, India
\item \Idef{org20}Department of Physics, Ohio State University, Columbus, Ohio, United States
\item \Idef{org21}Department of Physics, Sejong University, Seoul, South Korea
\item \Idef{org22}Department of Physics, University of Oslo, Oslo, Norway
\item \Idef{org23}Dipartimento di Elettrotecnica ed Elettronica del Politecnico, Bari, Italy
\item \Idef{org24}Dipartimento di Fisica dell'Universit\`{a} 'La Sapienza' and Sezione INFN Rome, Italy
\item \Idef{org25}Dipartimento di Fisica dell'Universit\`{a} and Sezione INFN, Cagliari, Italy
\item \Idef{org26}Dipartimento di Fisica dell'Universit\`{a} and Sezione INFN, Trieste, Italy
\item \Idef{org27}Dipartimento di Fisica dell'Universit\`{a} and Sezione INFN, Turin, Italy
\item \Idef{org28}Dipartimento di Fisica e Astronomia dell'Universit\`{a} and Sezione INFN, Bologna, Italy
\item \Idef{org29}Dipartimento di Fisica e Astronomia dell'Universit\`{a} and Sezione INFN, Catania, Italy
\item \Idef{org30}Dipartimento di Fisica e Astronomia dell'Universit\`{a} and Sezione INFN, Padova, Italy
\item \Idef{org31}Dipartimento di Fisica `E.R.~Caianiello' dell'Universit\`{a} and Gruppo Collegato INFN, Salerno, Italy
\item \Idef{org32}Dipartimento di Scienze e Innovazione Tecnologica dell'Universit\`{a} del  Piemonte Orientale and Gruppo Collegato INFN, Alessandria, Italy
\item \Idef{org33}Dipartimento Interateneo di Fisica `M.~Merlin' and Sezione INFN, Bari, Italy
\item \Idef{org34}Division of Experimental High Energy Physics, University of Lund, Lund, Sweden
\item \Idef{org35}Eberhard Karls Universit\"{a}t T\"{u}bingen, T\"{u}bingen, Germany
\item \Idef{org36}European Organization for Nuclear Research (CERN), Geneva, Switzerland
\item \Idef{org37}Excellence Cluster Universe, Technische Universit\"{a}t M\"{u}nchen, Munich, Germany
\item \Idef{org38}Faculty of Engineering, Bergen University College, Bergen, Norway
\item \Idef{org39}Faculty of Mathematics, Physics and Informatics, Comenius University, Bratislava, Slovakia
\item \Idef{org40}Faculty of Nuclear Sciences and Physical Engineering, Czech Technical University in Prague, Prague, Czech Republic
\item \Idef{org41}Faculty of Science, P.J.~\v{S}af\'{a}rik University, Ko\v{s}ice, Slovakia
\item \Idef{org42}Faculty of Technology, Buskerud and Vestfold University College, Vestfold, Norway
\item \Idef{org43}Frankfurt Institute for Advanced Studies, Johann Wolfgang Goethe-Universit\"{a}t Frankfurt, Frankfurt, Germany
\item \Idef{org44}Gangneung-Wonju National University, Gangneung, South Korea
\item \Idef{org45}Gauhati University, Department of Physics, Guwahati, India
\item \Idef{org46}Helsinki Institute of Physics (HIP), Helsinki, Finland
\item \Idef{org47}Hiroshima University, Hiroshima, Japan
\item \Idef{org48}Indian Institute of Technology Bombay (IIT), Mumbai, India
\item \Idef{org49}Indian Institute of Technology Indore, Indore (IITI), India
\item \Idef{org50}Inha University, Incheon, South Korea
\item \Idef{org51}Institut de Physique Nucl\'eaire d'Orsay (IPNO), Universit\'e Paris-Sud, CNRS-IN2P3, Orsay, France
\item \Idef{org52}Institut f\"{u}r Informatik, Johann Wolfgang Goethe-Universit\"{a}t Frankfurt, Frankfurt, Germany
\item \Idef{org53}Institut f\"{u}r Kernphysik, Johann Wolfgang Goethe-Universit\"{a}t Frankfurt, Frankfurt, Germany
\item \Idef{org54}Institut f\"{u}r Kernphysik, Westf\"{a}lische Wilhelms-Universit\"{a}t M\"{u}nster, M\"{u}nster, Germany
\item \Idef{org55}Institut Pluridisciplinaire Hubert Curien (IPHC), Universit\'{e} de Strasbourg, CNRS-IN2P3, Strasbourg, France
\item \Idef{org56}Institute for Nuclear Research, Academy of Sciences, Moscow, Russia
\item \Idef{org57}Institute for Subatomic Physics of Utrecht University, Utrecht, Netherlands
\item \Idef{org58}Institute for Theoretical and Experimental Physics, Moscow, Russia
\item \Idef{org59}Institute of Experimental Physics, Slovak Academy of Sciences, Ko\v{s}ice, Slovakia
\item \Idef{org60}Institute of Physics, Academy of Sciences of the Czech Republic, Prague, Czech Republic
\item \Idef{org61}Institute of Physics, Bhubaneswar, India
\item \Idef{org62}Institute of Space Science (ISS), Bucharest, Romania
\item \Idef{org63}Instituto de Ciencias Nucleares, Universidad Nacional Aut\'{o}noma de M\'{e}xico, Mexico City, Mexico
\item \Idef{org64}Instituto de F\'{\i}sica, Universidad Nacional Aut\'{o}noma de M\'{e}xico, Mexico City, Mexico
\item \Idef{org65}iThemba LABS, National Research Foundation, Somerset West, South Africa
\item \Idef{org66}Joint Institute for Nuclear Research (JINR), Dubna, Russia
\item \Idef{org67}Konkuk University, Seoul, South Korea
\item \Idef{org68}Korea Institute of Science and Technology Information, Daejeon, South Korea
\item \Idef{org69}KTO Karatay University, Konya, Turkey
\item \Idef{org70}Laboratoire de Physique Corpusculaire (LPC), Clermont Universit\'{e}, Universit\'{e} Blaise Pascal, CNRS--IN2P3, Clermont-Ferrand, France
\item \Idef{org71}Laboratoire de Physique Subatomique et de Cosmologie, Universit\'{e} Grenoble-Alpes, CNRS-IN2P3, Grenoble, France
\item \Idef{org72}Laboratori Nazionali di Frascati, INFN, Frascati, Italy
\item \Idef{org73}Laboratori Nazionali di Legnaro, INFN, Legnaro, Italy
\item \Idef{org74}Lawrence Berkeley National Laboratory, Berkeley, California, United States
\item \Idef{org75}Lawrence Livermore National Laboratory, Livermore, California, United States
\item \Idef{org76}Moscow Engineering Physics Institute, Moscow, Russia
\item \Idef{org77}National Centre for Nuclear Studies, Warsaw, Poland
\item \Idef{org78}National Institute for Physics and Nuclear Engineering, Bucharest, Romania
\item \Idef{org79}National Institute of Science Education and Research, Bhubaneswar, India
\item \Idef{org80}Niels Bohr Institute, University of Copenhagen, Copenhagen, Denmark
\item \Idef{org81}Nikhef, Nationaal instituut voor subatomaire fysica, Amsterdam, Netherlands
\item \Idef{org82}Nuclear Physics Group, STFC Daresbury Laboratory, Daresbury, United Kingdom
\item \Idef{org83}Nuclear Physics Institute, Academy of Sciences of the Czech Republic, \v{R}e\v{z} u Prahy, Czech Republic
\item \Idef{org84}Oak Ridge National Laboratory, Oak Ridge, Tennessee, United States
\item \Idef{org85}Petersburg Nuclear Physics Institute, Gatchina, Russia
\item \Idef{org86}Physics Department, Creighton University, Omaha, Nebraska, United States
\item \Idef{org87}Physics Department, Panjab University, Chandigarh, India
\item \Idef{org88}Physics Department, University of Athens, Athens, Greece
\item \Idef{org89}Physics Department, University of Cape Town, Cape Town, South Africa
\item \Idef{org90}Physics Department, University of Jammu, Jammu, India
\item \Idef{org91}Physics Department, University of Rajasthan, Jaipur, India
\item \Idef{org92}Physik Department, Technische Universit\"{a}t M\"{u}nchen, Munich, Germany
\item \Idef{org93}Physikalisches Institut, Ruprecht-Karls-Universit\"{a}t Heidelberg, Heidelberg, Germany
\item \Idef{org94}Politecnico di Torino, Turin, Italy
\item \Idef{org95}Purdue University, West Lafayette, Indiana, United States
\item \Idef{org96}Pusan National University, Pusan, South Korea
\item \Idef{org97}Research Division and ExtreMe Matter Institute EMMI, GSI Helmholtzzentrum f\"ur Schwerionenforschung, Darmstadt, Germany
\item \Idef{org98}Rudjer Bo\v{s}kovi\'{c} Institute, Zagreb, Croatia
\item \Idef{org99}Russian Federal Nuclear Center (VNIIEF), Sarov, Russia
\item \Idef{org100}Russian Research Centre Kurchatov Institute, Moscow, Russia
\item \Idef{org101}Saha Institute of Nuclear Physics, Kolkata, India
\item \Idef{org102}School of Physics and Astronomy, University of Birmingham, Birmingham, United Kingdom
\item \Idef{org103}Secci\'{o}n F\'{\i}sica, Departamento de Ciencias, Pontificia Universidad Cat\'{o}lica del Per\'{u}, Lima, Peru
\item \Idef{org104}Sezione INFN, Bari, Italy
\item \Idef{org105}Sezione INFN, Bologna, Italy
\item \Idef{org106}Sezione INFN, Cagliari, Italy
\item \Idef{org107}Sezione INFN, Catania, Italy
\item \Idef{org108}Sezione INFN, Padova, Italy
\item \Idef{org109}Sezione INFN, Rome, Italy
\item \Idef{org110}Sezione INFN, Trieste, Italy
\item \Idef{org111}Sezione INFN, Turin, Italy
\item \Idef{org112}SSC IHEP of NRC Kurchatov institute, Protvino, Russia
\item \Idef{org113}SUBATECH, Ecole des Mines de Nantes, Universit\'{e} de Nantes, CNRS-IN2P3, Nantes, France
\item \Idef{org114}Suranaree University of Technology, Nakhon Ratchasima, Thailand
\item \Idef{org115}Technical University of Split FESB, Split, Croatia
\item \Idef{org116}The Henryk Niewodniczanski Institute of Nuclear Physics, Polish Academy of Sciences, Cracow, Poland
\item \Idef{org117}The University of Texas at Austin, Physics Department, Austin, Texas, USA
\item \Idef{org118}Universidad Aut\'{o}noma de Sinaloa, Culiac\'{a}n, Mexico
\item \Idef{org119}Universidade de S\~{a}o Paulo (USP), S\~{a}o Paulo, Brazil
\item \Idef{org120}Universidade Estadual de Campinas (UNICAMP), Campinas, Brazil
\item \Idef{org121}University of Houston, Houston, Texas, United States
\item \Idef{org122}University of Jyv\"{a}skyl\"{a}, Jyv\"{a}skyl\"{a}, Finland
\item \Idef{org123}University of Liverpool, Liverpool, United Kingdom
\item \Idef{org124}University of Tennessee, Knoxville, Tennessee, United States
\item \Idef{org125}University of the Witwatersrand, Johannesburg, South Africa
\item \Idef{org126}University of Tokyo, Tokyo, Japan
\item \Idef{org127}University of Tsukuba, Tsukuba, Japan
\item \Idef{org128}University of Zagreb, Zagreb, Croatia
\item \Idef{org129}Universit\'{e} de Lyon, Universit\'{e} Lyon 1, CNRS/IN2P3, IPN-Lyon, Villeurbanne, France
\item \Idef{org130}V.~Fock Institute for Physics, St. Petersburg State University, St. Petersburg, Russia
\item \Idef{org131}Variable Energy Cyclotron Centre, Kolkata, India
\item \Idef{org132}Vin\v{c}a Institute of Nuclear Sciences, Belgrade, Serbia
\item \Idef{org133}Warsaw University of Technology, Warsaw, Poland
\item \Idef{org134}Wayne State University, Detroit, Michigan, United States
\item \Idef{org135}Wigner Research Centre for Physics, Hungarian Academy of Sciences, Budapest, Hungary
\item \Idef{org136}Yale University, New Haven, Connecticut, United States
\item \Idef{org137}Yonsei University, Seoul, South Korea
\item \Idef{org138}Zentrum f\"{u}r Technologietransfer und Telekommunikation (ZTT), Fachhochschule Worms, Worms, Germany
\end{Authlist}
\endgroup

\end{document}